\documentclass[pra,twocolumn,showpacs,preprintnumbers,amsmath,amssymb]{revtex4}

\usepackage{graphicx}
\usepackage{dcolumn}
\usepackage{bm}

\begin{document}

\preprint{APS/123-QED}

\title{A non-adiabatic approach to entanglement distribution over long distances}

\author{Mohsen Razavi}
\email{mora158@mit.edu}
\altaffiliation[Now with the ]{Institute for Quantum Computing, University of Waterloo, Waterloo, ON, Canada N2L 3G1.}
\author{Jeffrey H. Shapiro}
 \affiliation{Research Laboratory of Electronics\\
Massachusetts Institute of Technology\\
Cambridge, Massachusetts 02139 USA}

\date{\today}

\begin{abstract}
Entanglement distribution between trapped-atom quantum memories, viz. single atoms in optical cavities, is addressed. In most scenarios, the rate of entanglement distribution depends on the efficiency with which the state of traveling single photons can be transferred to trapped atoms. This loading efficiency is analytically studied for two-level, $V$-level, $\Lambda$-level, and double-$\Lambda$-level atomic configurations by means of a system-reservoir approach. An off-resonant non-adiabatic approach to loading $\Lambda$-level trapped-atom memories is proposed, and the ensuing trade-offs between the atom-light coupling rate and input photon bandwidth for achieving a high loading probability are identified. The non-adiabatic approach allows a broad class of optical sources to be used, and in some cases it provides a higher system throughput than what can be achieved by adiabatic loading mechanisms. The analysis is extended to the case of two double-$\Lambda$ trapped-atom memories illuminated by a polarization-entangled biphoton.
\end{abstract}


\maketitle

\section{INTRODUCTION}
\label{sect:intro}
The future  of quantum information science---which holds the promise of revolutionary improvements in computation \cite{Shor94, Gro96}, secure communication \cite{BB84}, and precision measurement \cite{prec_meas}---depends on our ability to develop its demanding technology. Some of the required building blocks have already been developed, e.g.,  first generation  quantum key distribution systems are now commercially available \cite{commercial}. With recent advances in single-photon technology---the development of high-flux sources of entangled photons \cite{chris, ultrabright} as well as efficient (photon-number resolving) single-photon detectors \cite{SPD1,SPD2,SPD3}---it is foreseeable that the next generation of secure long-distance communication systems will be within reach in a few years. The key element, which is still missing from such systems, is a reliable quantum memory that can store and manipulate quantum information. Whereas trapped-ion systems are promising choices for quantum processing, the lack of an efficient optical interface hinders their being used in long-distance quantum communication systems. Some neutral alkali atoms, however, e.g. rubidium and cesium, are appropriate choices for capturing the states of single photons and storing them in their metastable hyperfine levels. In order to enhance the interaction of single photons and atomic systems, we have to either trap single atoms in  high-finesse optical cavities \cite{trap, trapatom1, trapatom2, trapatom3} or use ensembles of atoms \cite{DLCZ,cohtime2, kuz2, Kuz1,Kuz3, cohtime3}. In either case, it is an interesting problem to quantify how efficiently we can transfer the state of a single photon  to an atomic quantum memory, or, in other words, how efficiently we can {\em load} a quantum memory with a desired state. In this paper, we study the loading problem in the context of trapped-atom quantum memories, viz. single atoms trapped in high-$Q$ optical cavities, and find analytical solutions for the loading probability in different scenarios in which either a single photon or a pair of entangled photons are driving the quantum memory units. That analysis results in a novel non-adiabatic approach to loading quantum memories with $\Lambda$ and double-$\Lambda$  atomic configurations. We also  investigate the effect of various system parameters  on the loading probability, including the atom-light coupling rate, the input light bandwidth, the cavity/atomic decay rate, and timing offsets.    
Atomic-ensemble memories will not be explicitly addressed in this paper, however, most of the analytical results that we obtain are directly applicable to such systems. We elaborate more on this issue in the discussion section. 

Entanglement is a quantum resource by which two parties share a joint state that cannot be written in a tensor-product form. This state provides a stronger-than-classical correlation between two quantum systems by which one can perfectly infer quantum measurement results made on one system by observing the results of quantum measurements made on the other. This quantum correlation is behind various  applications in quantum communication \cite{Bennett} and distributed quantum processing \cite{Ray1}. In teleportation, entanglement serves as a quantum wire. Once established, it can be used for one-time communication between the transmitter and the receiver. In distributed quantum processors, entanglement serves as a computational resource. The main trick in both of these applications is that we have connected our parties, in advance of any subsequent actions, via entanglement. This way, we need not worry about the channel loss or physical transportation of a qubit once we have established entanglement. We ought to worry, however, about how to generate, distribute, and maintain entanglement over long distances, i.e., distances over which our  physical system undergoes decoherence. 

\begin{figure}
\centering
\includegraphics [width=\linewidth] {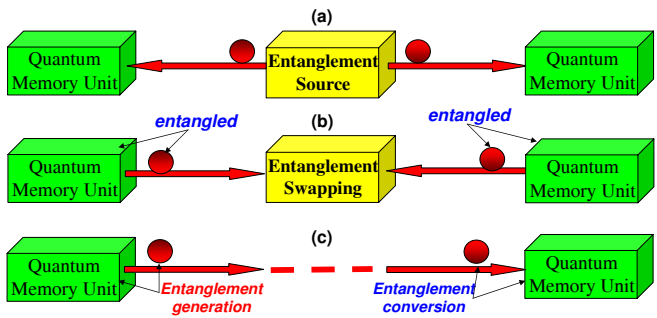}
\caption{\label{telepapp}
(Color online) Three architectures for entanglement distribution: (a) To-the-memory architecture, in which two entangled photons, generated at the source, carry the entangled state to the quantum memory units and load them with this state. (b) From-the-memory architecture, in which, by the use of a common source, each memory generates a flying qubit entangled with itself. Entanglement swapping is then accomplished by means of a Bell-state measurement on the photons, at the midpoint, which leaves the memories in an entangled state. (c) Memory-to-memory configuration, in which a flying qubit entangled with one of the memories is generated that then propagates to and loads the other memory, i.e., transfers its state to that memory.}
\end{figure}

There are three commonly suggested approaches for long-distance entanglement distribution, all of which use photons for the transmission of quantum information and atomic systems for its storage. They are shown schematically in Fig.~\ref{telepapp}. In the first approach, we produce a pair of entangled photons at an optical source, and let the photons travel to and be stored in the quantum memory units \cite{MIT/NU}. 
In the second approach, we use  entanglement swapping \cite{swap} to distribute the entanglement.  We first generate entanglement between a standing qubit and a flying qubit at  each end of Fig.~\ref{telepapp}(b). We then let the photons propagate  to the midpoint between memories where a Bell-state measurement  apparatus destroys them, leaving the quantum memory units in an entangled state \cite{DLCZ, DK03}. The last architecture, shown in Fig.~\ref{telepapp}(c),  is a hybrid of the previous ones. In this approach, we generate a single photon entangled with one of the quantum memories, and let this photon propagate to and load the other memory, i.e., transfer its state to that memory \cite{Kuz3,nontelep}.

In all of these entanglement distribution schemes, a figure of merit for system performance is its throughput, viz. how often we obtain an entangled pair of quantum memories. Throughput is a function of different system parameters. For instance, in the to-the-memory architecture, Fig.~\ref{telepapp}(a), it depends on the source generation rate of entangled photons, the path and coupling losses, and the loading efficiency. System loss has the unavoidable effect of reducing the throughput. Loss, however, does not affect the source generation rate or the nature of interaction between a single photon and a single atom. Moreover, for trapped-atom quantum memories, there are nondestructive loading verification techniques that allow us to detect  photon loss \cite{trap}. Hence, throughout the paper, we assume that the system is lossless, and we focus on the interaction between the source output and the memory modules. 

The interaction between a trapped atom and light is governed by several factors, among them are the input photon bandwidth, the cavity decay rate, and the atom-light coupling rate. In order to realize a strong atom-light interaction, the photons must have bandwidths comparable to atomic linewidths, which are typically on the order of tens of MHz. Higher optical bandwidths  may afford increased source generation rates---as is the case for spontaneous parametric downconversion sources whose bandwidths are on the order of THz \cite{PDC1,PDC2,PDC3,PDC4}---and, hence, increased total throughput. However, if  higher source bandwidth comes with  lower spectral brightness, throughput may be decreased, owing to a decrease in the loading probability. It may be possible to maintain a high loading probability by employing a higher atom-light coupling rate. This last approach requires a more demanding implementation, however, because, for trapped-atom quantum memories, the coupling rate is increased by reducing the cavity length. All these issues necessitate developing a quantitative loading analysis that quantifies the interplay between the input photons' bandwidths and the quantum memory parameters, i.e., cavity decay rate and atom-light coupling rate. This paper seeks to answer these questions, which are of practical importance in designing quantum memory systems, by quantifying the trade-offs inherent in adiabatic and non-adiabatic loading mechanisms.

Another important issue in dealing with the interaction of trapped atoms and light is decoherence. There are a variety of mechanisms by which our system may decohere. Decay of photons out of the cavity and atomic spontaneous emission are the two factors that will be considered in our analysis. The latter effect can be alleviated by using off-resonant transitions, which will be employed in many scenarios in this paper. There are other sources of decoherence---some related to the employed trapping and cooling schemes \cite{dephasing}, and some due to the  oversimplified picture that we use for atomic systems \cite{cohtime}---which will not be addressed in our paper.

The rest of this paper is organized as follows. In Section~\ref{Secsingle}, we study the loading process for single trapped-atom quantum memories that are illuminated by single photons. The methodology that we employ in this section will be applicable to systems that use the memory-to-memory configuration for entanglement distribution.  We start with the basic case of a two-level atom, and then extend our results to $\Lambda$- and double-$\Lambda$-level atoms. We propose a  non-adiabatic loading mechanism for the latter cases, and compare it to the previously proposed adiabatic loading techniques \cite{nontelep, fleisch}. In Section~\ref{Secentangle} we treat an example of the to-the-memory configuration in which the source output is a polarization-entangled biphoton, and the trapped atoms have double-$\Lambda$ configurations. This is a close approximation to the architecture proposed by researchers from the Massachusetts Institute of Technology and Northwestern university (termed MIT-NU hereafter) \cite{trap,MIT/NU,Brent}. Thus the results presented here comprise the first loading analysis of the MIT-NU system that accounts for the intracavity atoms.

\section{Trapped-atom Quantum memories driven by single photons} 
\label{Secsingle}

In this section, we evaluate the loading probability for trapped-atom quantum memories, viz. single atoms in high-finesse optical cavities, driven by single photons. The loading protocol is successful if we transfer the photon state to the corresponding atomic levels. We do not deal with  the complexity associated with trapping neutral atoms, and assume that the single atoms can be trapped at a fixed point in the cavity for a time sufficient for loading to occur \cite{trapatom1}. Throughout the paper, optical cavities are assumed to be lossless and single ended. The external photon illuminates the cavity through its partially reflecting mirror in a specific spatial and polarization mode matched to the cavity's mode of interest. We assume that the incoming light has such a narrow bandwidth that it only interacts with a single longitudinal mode of the cavity. To model the temporal content of the incoming photon,  we consider a reservoir of harmonic oscillators corresponding to a continuum of modes, as suggested in \cite{GC}. This is also in accord with the quantum representation of  traveling-wave light \cite{Loudon}. We then employ a system-reservoir approach to analyze the loading dynamics.

\subsection{Two-level Trapped Atoms}
\label{Sectwolevel}

\begin{figure}
	\centering
		\includegraphics  [width = \linewidth] {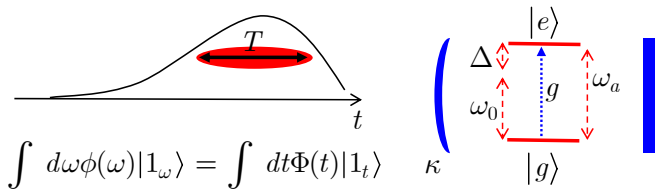}
	\caption{(Color online) A single two-level atom trapped in a single-ended high-finesse optical cavity. A single photon illuminates the cavity at the proper frequency to excite the atom to its upper state.}
	\label{twolevel}
\end{figure}

Suppose that there is a single two-level atom, with excited state $\left| e \right\rangle $ and ground state $\left| g \right\rangle $, inside a single-ended high-$Q$ cavity with decay rate $\kappa$. We assume that the frequency-$\omega_a$ atomic transition between $\left| e \right\rangle $ and $\left| g \right\rangle $ is coupled, at coupling rate $g$, to the cavity field operator $\hat b$ of frequency $\omega_0$. This implies a detuning $\Delta = \omega_0 - \omega_a$, which is assumed to be much less than $\omega_0$. In general, $g$ and $\kappa$ can both be functions of time. Here, we assume that $\kappa$ is a constant. Suppose that this trapped-atom module, with the atom initially in its ground state and no photon in the cavity, is illuminated by a single photon, as shown in Fig.~\ref{twolevel}. This driving source can be modeled by a set of annihilation operators $\hat a_\omega$, each corresponding to a different temporal (spectral) mode of frequency $\omega$, that satisfy $[\hat a_\omega, \hat a^ \dag _{\omega'}]=\delta (\omega - \omega')$ \cite{Loudon}. We assume that the driving field is initially in the following single-photon state
\begin{eqnarray}
\label{initial_reservoir_state}
\left| \psi_0 \right\rangle & = & \int {d\omega \phi(\omega) \left| 1_\omega \right\rangle} \nonumber \\
& = &  \int {dt \Phi(t) \left| 1_t \right\rangle},
\end{eqnarray}
where $\int {d\omega |\phi(\omega)|^2 } = 1$, and $| 1_\omega \rangle = \hat a_\omega^\dag |{\bf 0}\rangle_R$ is the multi-mode state representing one photon at frequency $\omega$ and the vacuum state for all other modes. Here,  $|{\bf 0}\rangle_R$ is the reservoir's multi-mode vacuum state. We can equivalently describe the initial state in the time domain by introducing the temporal pulse shape $ \Phi(t) =   \int {d\omega e^{-i \omega t} \phi(\omega) / {\sqrt{2\pi}}} $ associated with the incoming photon, and single-photon states $ |1_t\rangle =   \int {d\omega e^{i \omega t} |1_\omega \rangle / {\sqrt{2\pi}}}$ associated with time $t$, as shown in Eq.~(\ref{initial_reservoir_state}). Here and elsewhere in this paper, we assume that the initial time is before  any possible interaction between the source and the memory. For simplicity, and without loss of generality, we assume that this initial time is $0$. We will relax this assumption after obtaining our final results. 

The Hamiltonian for the above system is given by \cite{SZ}
\begin{equation}
\hat H_a = \hat H_{cc} + \hbar \omega_a \hat \sigma_{ee} + \hbar g (\hat b^\dag \hat \sigma_{ge} + \hat b \hat \sigma_{eg}),
  \end{equation}
where $\hat \sigma_{ij} = \left| i \right\rangle \left\langle j \right|$, $i,j \in \{g,e\}$, and
\begin{equation}
\label{Ham2L_Bath}
\hat H_{cc}  =  \hbar \int {d\omega \, \omega \hat a_\omega^\dag  \hat a_\omega} + \hbar \omega_0 \hat b^\dag \hat b + \hbar \Gamma \int {d\omega (\hat a_\omega^\dag \hat b + \hat b^\dag \hat a_\omega)} ,
\end {equation}
with $\Gamma \equiv \sqrt{\kappa / \pi}$ being the coupling constant that connects the external world to the cavity  \cite{GC}. Because of the source's narrow bandwidth, we can and will assume that all integrals in Eq.~(\ref{Ham2L_Bath}) run from $-\infty$ to $+\infty$.

Neglecting for now the decoherence mechanisms that  influence the time evolution, the quantum system consisting of the cavity, the atom, and the reservoir is closed, with exactly one excitation at any time $t$. A general quantum state of the system at time $t$ can thus be written as
\begin{eqnarray}
\label{qstate2}
|\psi (t) \rangle &=&  \int {d\omega \alpha_\omega(t) | 1_\omega \rangle | 0 \rangle_b | g \rangle } + e^{-i \omega_0 t}\beta(t) | {\bf 0} \rangle_R | 1 \rangle_b | g \rangle \nonumber\\
&& + e^{-i \omega_0 t} c_e (t) | {\bf 0} \rangle_R | 0 \rangle_b | e \rangle ,
\end {eqnarray}
where $| k \rangle_b$ represents the $k$-photon Fock state of the cavity mode. 

The goal of this section is to provide analytical results for the memory loading probability---the probability of absorbing the external photon by the trapped atom---$|c_e(t)|^2$. The atom in our system is connected to the external single photon via the cavity mode. The interaction of the reservoir and the cavity mode can be analyzed by applying the Schr\"odinger equation, $i \hbar|\dot \psi (t) \rangle = \hat H_{a} |\psi (t) \rangle$, to Eq.~(\ref{qstate2}), to obtain
\begin{subequations}
\label{dotalphabeta}
\begin{eqnarray}
\label{dotalphaomega}
& \dot \alpha_\omega (t) = -i ( \omega \alpha_\omega (t) + \Gamma e^{-i \omega_0 t}  \beta(t) ) ,& \\
\label{dotbetat1}
& \dot \beta(t) = -i  \Gamma e^{i \omega_0 t} \int {d \omega \, \alpha_\omega (t)}. &  
\end{eqnarray}
\end{subequations}
Solving for $\alpha_\omega(t)$ in Eq.~(\ref{dotalphaomega}), and then plugging the result into Eq.~(\ref{dotbetat1}), we obtain
\begin{equation}
\label{dotbetat2}
\dot \beta (t) =  - i g c_e(t) -i \sqrt{2 \kappa}  \Phi_b(t) - \kappa \beta (t),
\end{equation}
where $\Phi_b(t) = e^{  i \omega_0 t} \Phi(t)$ is the baseband input pulse shape provided that the center frequency of the incoming light is $\omega_0$. The second term on the right-hand side of the above equation accounts for the effect of the input photon. The last term accounts for the decay of the photon out of the cavity, and in essence, is similar to the decay term in the Weisskopf-Wigner theory of spontaneous emission \cite{SZ}, in which a two-level system initially in its excited state interacts with a reservoir initially in its vacuum state. In this way, we can incorporate the overall spontaneous decay, with rate $\gamma$, from the excited state of the atom to the modes other than the cavity's mode of interest, and thus obtain
\begin{equation}
\label{dotce}
\dot c_e (t) =  i \Delta  c_e (t) - i g \beta(t)  - \gamma c_e (t) ,
\end{equation}
where the first two terms on the right-hand side of Eq.~(\ref{dotce}) represent the coherent evolution of $c_e(t)$ in accord with the Schr\"odinger equation $i \hbar|\dot \psi (t) \rangle = \hat H_{a} |\psi (t) \rangle$. 

Equations~(\ref{dotbetat2}) and (\ref{dotce}) model the loading dynamics of a two-level trapped-atom memory illuminated by a single photon with pulse shape $\Phi(t)$. For the general case of time-dependent system parameters, these equations can be solved numerically. 
For the special case of time-independent $g$, $\kappa$, $\gamma$, and $\Delta$, we can solve the above equations analytically using the Laplace-transform techniques, which yield 
\begin{equation}
\label{betatfinal}
 \beta(t) = \frac{- i \sqrt{2 \kappa}}{\xi}  \int_0^t { d \tau  \Phi_b (\tau) (\kappa'_+ e^{-\kappa_+ (t- \tau)} - \kappa'_- e^{ - \kappa_- (t- \tau)})}   \end{equation}
and
\begin{equation}
\label{cet}
 c_e(t) =  \frac{g \sqrt{2 \kappa}}{ \xi} \int_0^t { d \tau  \Phi_b (\tau) (e^{-\kappa_+ (t- \tau)} - e^{ - \kappa_- (t- \tau)})} , 
\end{equation}
where
 \begin{eqnarray}
 \label{loadingparam} 
 & \kappa_\pm = (\kappa + \gamma' \pm \xi)/2, \kappa'_\pm = \kappa_\pm -\gamma',& \nonumber \\
& \xi = \sqrt{(\kappa-\gamma')^2 - 4 g^2 },\gamma' = \gamma - i \Delta .&
\end{eqnarray}
The lower limit zero in the above integrals represents the initial time of interaction. It can be changed  to accommodate the input pulses that start before $t=0$ as well.

Note that Eqs.~(\ref{dotbetat2}) and (\ref{dotce}), with initial values $c_e(0)=\beta(0)=0$, are  linear in response to the input pulse shape. For instance, assuming that  $\Phi_b(t) = \Phi_1(t) + \Phi_2(t)$,  $\beta(t)$ will be equal to $\beta_1 (t) + \beta_2 (t)$, where $\beta_i(t)$, for $i=1,2$, is given by Eq.~(\ref{betatfinal}) with $\Phi_b(t)$ replaced by $\Phi_i(t)$. Similarly, we can think of the initial state in Eq.~(\ref{initial_reservoir_state}) as a superposition of its infinitesimal constituents $d\omega \phi(\omega) |1_\omega\rangle$,  to which, from Eq.(\ref{cet}), we can associate slowly-varying  probability amplitudes $c_{e,\omega}(t)$ given by
\begin{eqnarray}
\label{ceomegat}
\lefteqn{c_{e,\omega}(t) d \omega =  (g \sqrt{ \kappa / \pi} / \xi ) \times}  \nonumber \\ 
& \int_0^t { d \tau  d \omega \phi(\omega)e^{-i(\omega-\omega_0)\tau} (e^{-\kappa_+ (t- \tau)} - e^{ - \kappa_- (t- \tau)})} .&
\end{eqnarray}
We then have $c_e(t) = \int{d \omega c_{e,\omega}(t)}$ as obtained in Eq.~(\ref{cet}). We use this result later in Section~\ref{Secentangle}.

The system-reservoir calculation  gives a more compact form for the loading probability, $|c_e (t)|^ 2$, than what we  obtained in \cite{QELS} via the Heisenberg-Langevin equations \cite {GC} for the case of on-resonance illumination at $\gamma =0 $. The two results are nevertheless equivalent, and their equivalence can be verified by a tedious algebraic manipulation that we shall omit. 

\subsubsection{Numerical results}
\label{SubsecNonadbNumres}

There are several physical parameters that are of practical importance  in loading a trapped-atom quantum memory: cavity decay rate, spontaneous decay rate, atom-light coupling rate, and input bandwidth.  However, from Eq.~(\ref{cet}), it can be seen that the loading probability is governed by the ratios of these parameters. Thus, for $\Delta = 0$, Eq.~(\ref{cet}) can be written in terms of the dimensionless parameters $\kappa T$, $g / \kappa$, $\gamma / g$, and $t/T$, as follows
\begin{eqnarray}
\hspace{-.14in}
c_e (t) & = & \frac{\sqrt {2 \kappa T}} {\xi / g} 
\int_0^{t/T} { du \Phi'_b(u) \times}
\nonumber \\
&& \left[ e^{ - \kappa T (\kappa_+ / \kappa)(t/T - u) }  
  -  e^{ - \kappa T ( \kappa_- / \kappa)(t/T - u) } \right] ,
\end{eqnarray}
where $T$ is the effective width of $\Phi_b(t)$, and $\Phi_b(t) = \Phi'_b(t/T)/\sqrt{T} $, i.e., $\Phi'_b(u)$ is a compressed/stretched version of  $\Phi_b(t)$, which has unity width, in its normalized time coordinate, and has been normalized to satisfy $\int {d u |\Phi'_b(u)|^2  } = 1$.

\begin{figure}
	\centering
		\includegraphics  [width = 0.7\linewidth] {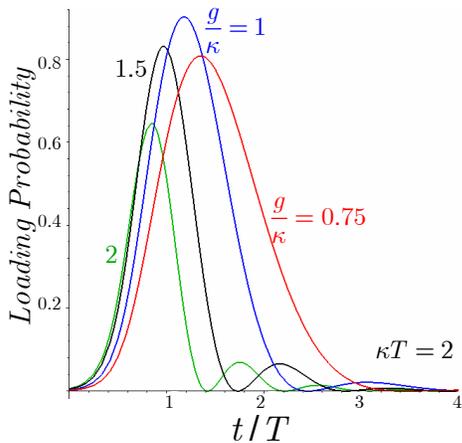}
	\caption{(Color online) Loading probability versus normalized time for $\kappa T = 2$ and a hyperbolic secant  pulse shape at $\gamma=\Delta =0$. Upon arrival of the photon, there is a peak in the loading probability, which occurs at the time that the photon is most likely to have absorbed by the atom. This maximum loading probability is a function of $g / \kappa$, and there exists an optimum value for this parameter that maximizes the chance of loading.}
	\label{optgk}
\end{figure}

For a constant value of $g$, what we expect from the atom-photon interaction is a damped Rabi oscillation. The incoming photon is initially transferred to the cavity mode, which interacts with the atom and drives a $|g\rangle$-to-$|e\rangle$ transition. That will be followed by the reverse transition which releases a photon into the cavity. These oscillations continue until the photon leaks out of the cavity or the atom spontaneously emits a photon into a non-cavity mode. 
Figure~\ref{optgk} shows the loading probability for the two-level atom of Fig.~\ref{twolevel} in the ideal case of $\gamma = \Delta =0$ for a hyperbolic secant  pulse shape $\Phi'_b(u) = \sqrt{2}{\rm sech}[4(u-1)]$. In this figure, we have fixed the value of $\kappa T$ to $2$ and varied the value of $g/ \kappa$. Counterintuitively, it can be seen that increasing the coupling rate $g$ does not necessarily improve the loading probability, but there exists an optimum value of $g/\kappa$, which maximizes the loading probability. Higher ratios than this optimum value just increase the number of Rabi flops between the two atomic levels, making it harder to find the atom in its excited state with high probability. However, by this method we cannot hold the atom in its excited state unless we drive the value of $g$ to zero at an appropriate point in the process, $t=T_{Load}$, when  the loading probability attains its maximum. In the next section, we show how we can turn off the coupling  in a $\Lambda$-level atom, in which we can use a control field to vary $g$.

\begin{figure}
	\centering
		\includegraphics  [width = 0.7\linewidth] {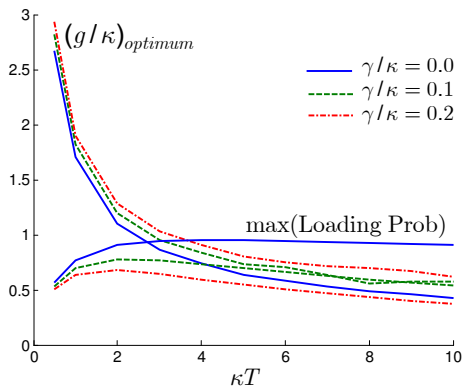}
	\caption{(Color online) The optimum coupling rate, and the maximum loading probability achieved at this rate, versus $\kappa T$ for loading a two-level atom by a single photon with hyperbolic secant  pulse shape on resonance. Several spontaneous decay rates, $\gamma$, are considered.}
	\label{designp3}
\end{figure}

Figure~\ref{designp3} shows the optimum value of $g/\kappa$ versus $\kappa T$ for several values of $\gamma$ and $\Delta =0$. Several things are noticeable in this figure. First, for a fixed $\kappa$, it can be seen that the required value for the coupling rate will decrease as we increase the pulse width, or equivalently, when our source's bandwidth decreases. The decrease that we achieve, however, by doing so is exponentially smaller as we go to larger values of $\kappa T$. For instance (not shown on the graph), to get $(g/\kappa)_{optimum} = 0.1$, we need $\kappa T$ to be approximately 100. It can also be seen that the optimum coupling rate will not change significantly in the presence of decoherence. Nonzero spontaneous decay rates, however, will reduce the maximum loading probability---which could be above 90\% otherwise---as shown in the lower part of Fig.~\ref{designp3}. Longer pulses, which require  longer loading times, are more exposed to the decay process, and so, their maximum loading probabilities are lower. Spontaneous decay can be alleviated by using large enough detunings in a $\Lambda$-level atomic system.

\begin{figure}
	\centering
		\includegraphics  [width = \linewidth] {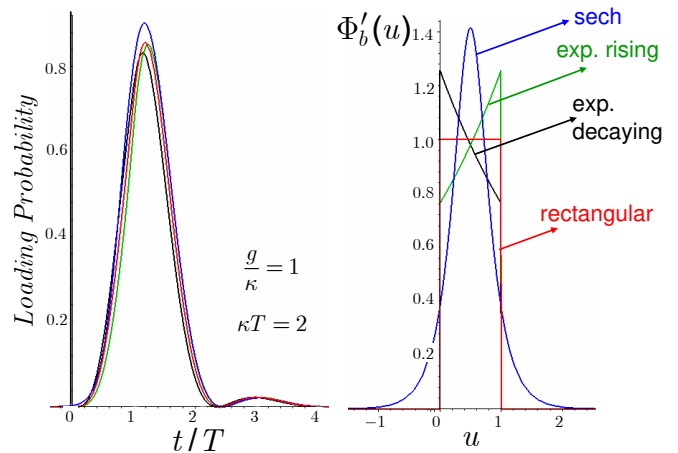}
	\caption{(Color online) Left: loading probability at $\kappa T = 2$, $g / \kappa =1$, and $\gamma=\Delta =0$ for various pulse shapes. Right: the baseband pulse shapes. In order of their maximum loading probabilities they are: the hyperbolic secant  (with the highest peak), rectangular, exponentially-rising, and exponentially-decaying pulse shapes.}
	\label{pshape}
\end{figure}

Figure~\ref{pshape} shows the loading probability as a function of time for $g/\kappa=1$ and $\kappa T=2$, for different pulse shapes of approximately the same effective width and $\gamma = \Delta  = 0$. The four different shapes we have used are: a hyperbolic secant  pulse; a rectangular pulse; an exponentially-rising pulse; and an exponentially-decaying pulse. The difference between their loading probabilities is seen to be very minor. It seems that having matched the input bandwidth to the cavity parameters $g$ and $\kappa$, we can achieve high loading probability regardless of the photon pulse shape. For the rest of this paper, we only consider the hyperbolic secant  pulse shape in our calculations.

\subsection{$\Lambda$-level Trapped Atoms}
\label{Seclambda}

The two-level atom that we analyzed in the previous section, while the easiest to study, is not a practical solution for quantum storage. For long lifetime storage, we have to store the quantum information in metastable atomic states that have low energy levels as well as low decay rates, e.g. the hyperfine levels in the $5\,{}^2 S_{1/2}$ orbital of the rubidium atom. Whereas the direct transition between such metastable levels are usually dipole forbidden,  we can connect these states, via a third excited level, by inducing  a Raman transition driven by a single photon on one leg and a classical control field on the other. The spontaneous decay from the upper state can then be mitigated by including large detunings. The resulting $\Lambda$-level atom  is at the core of all interesting neutral-atom quantum memory units. This section studies the loading problem in the context of trapped-atom memories with $\Lambda$-level configurations.

\begin{figure}
	\centering
		\includegraphics  [width = \linewidth] {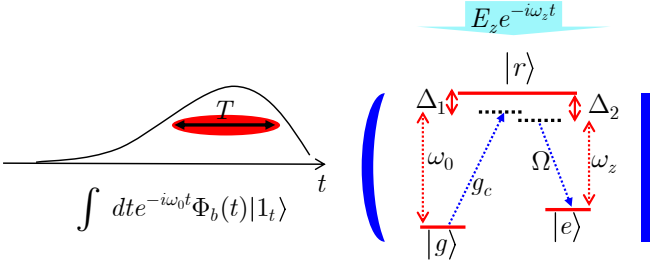}
	\caption{(Color online) A $\Lambda$-level trapped atom illuminated by a single photon. The cavity mode corresponding  to the single photon drives the atom from its ground state $|g\rangle$ to the auxiliary state $|r\rangle$. A $z$-polarized beam then shelves the atom in the metastable level $|e\rangle$. Here, $g_c$ is the vacuum Rabi frequency for the $|g\rangle$-to-$|r\rangle$ transition, and $\Omega$ is the Rabi frequency associated with the control field, which is proportional to $|E_z|$. These transitions are off-resonant by detunings $\Delta_1$ and $\Delta_2$, respectively, as defined in the text. The control field may also include a time-varying phase $\phi_z(t)$.}
	\label{lambdalevel} 
\end{figure}

Figure~\ref{lambdalevel} shows a trapped $\Lambda$-level atom and its corresponding driving beams. Here, we assume that an external single-photon beam, in the initial state given by Eq.~(\ref{initial_reservoir_state}), is spatially matched to a cavity mode with annihilation operator $\hat b$ and resonance frequency $\omega_0$. This cavity mode drives the $|g\rangle$-to-$|r\rangle$ transition with coupling rate $g_c$. There may exist a detuning $\Delta_1 \equiv \omega_0 - \omega_{gr}$ in this transition, where $\omega_{ij}$ denotes the transition frequency between levels $|i\rangle$ and $|j\rangle$, for $i,j \in \{g,r,e\}$. The second beam is assumed to be a $z$-polarized classical wave with frequency $\omega_z$ and a possibly time-dependent phase $\phi_z(t)$. This field's amplitude, which is under our control and may vary with time, determines the Rabi frequency $\Omega(t)$. It couples $|r\rangle$ and $|e\rangle$ via the following Hamiltonian, which is obtained under the rotating-wave approximation \cite{SZ}:
\begin{eqnarray}
\label{HamR}
\hat H_{b} &=& \hat H_{cc} + \hbar \omega_{gr} \hat \sigma_{rr} + \hbar \omega_{ge} \hat \sigma_{ee} + \hbar g_c (\hat b^\dag \hat \sigma_{gr} + \hat \sigma_{rg} \hat b) \nonumber \\
& +& \hbar \Omega(t) (e^{-i [\omega_z t+\phi_z(t)]} \hat \sigma_{re} + e^{i [\omega_z t + \phi_z(t)]} \hat \sigma_{er}),
\end{eqnarray}
where $\Omega(t)$ and $\phi_z(t)$ are assumed to be real. The $|r\rangle$-to-$|e\rangle$ transition may also be off-resonant, by a detuning $\Delta_2 \equiv \omega_z - \omega_{er}$. Similar to the previous section, neglecting all decoherence mechanisms, with no initial excitation in the cavity and assuming that the atom is initially in its ground state $|g\rangle$, the most general time evolution for the state of the system takes the following form
\begin{eqnarray}
\label{lambda_ini_state}
|\psi(t) \rangle & = & \int {d\omega \alpha_\omega(t) | 1_\omega \rangle | 0 \rangle_b | g \rangle} \nonumber \\
& + & e^{- i \omega_0 t} \beta(t) |G\rangle 
 +  e^{- i \omega_0 t} c_r(t) |R\rangle \nonumber \\
& + & e^{-i (\omega_0-\omega_z) t + i \phi_z(t)} c_e(t) |E \rangle,
\end{eqnarray}
where
\begin{subequations}
\begin{eqnarray}
& |G\rangle \equiv | {\bf 0} \rangle_R | 1 \rangle_b | g \rangle ,&  \\
& |R \rangle \equiv | {\bf 0} \rangle_R | 0 \rangle_b | r \rangle ,& \\
& |E \rangle \equiv | {\bf 0} \rangle_R | 0 \rangle_b | e \rangle .&
\end{eqnarray}
\end{subequations}
Applying the Schr\"odinger equation  $i \hbar|\dot \psi (t) \rangle = \hat H_{b} |\psi (t) \rangle$ to Eq.~(\ref{lambda_ini_state}), we get
\begin{subequations}
\label{lamsch}
\begin{eqnarray}
& \hspace{-0.24in}
\dot \beta(t) = -i g_c c_r(t) -i \sqrt{2 \kappa} \Phi_b(t) - \kappa \beta(t) &  \\
& \hspace{-0.24in}
\dot c_r (t) = -i g_c \beta (t) + i \Delta_1 c_r (t) - i \Omega (t) c_e (t) - \gamma_r c_r(t)&  \\
& \hspace{-0.24in}
\dot c_e (t) = - i \Omega(t) c_r (t) - i (\Delta_2 -\Delta_1 + \dot \phi_z(t)) c_e (t), &
\end{eqnarray}
\end{subequations}
where we used Eq.~(\ref{dotbetat2}) in the first equation and included a non-cavity decay rate $\gamma_r$ for the upper state. We assume that the corresponding decay rate to the state $|e\rangle$ is negligible for the purpose of loading; it comes into play if we need to determine the storage time of the quantum memory, which is not a topic of interest in this paper. 
For large enough detunings, i.e., $\Delta_1,\Delta_2 \gg \Omega, g_c$, we can adiabatically eliminate the upper state $|R\rangle$ by assuming that $\dot c_r (t) =0$. As a result, we obtain
\begin{equation}
c_r(t) = \frac{g_c/\Delta_1}{ \Gamma_r} \beta(t) + \frac{\Omega(t)/\Delta_1}{\Gamma_r} c_e(t) 
\end{equation}
where $\Gamma_r \equiv 1+ i \gamma_r/\Delta_1$. Plugging the above equation into Eq.~(\ref{lamsch}), we then obtain
\begin{subequations}
\label{schadbelm}
\begin{eqnarray}
 \dot \beta (t) & = & - i \frac{g_c^2 / \Delta_1}{\Gamma_r} \beta (t) - i \frac{g(t)}{\Gamma_r} c_e (t) \nonumber \\
 && - i \sqrt{2 \kappa} \Phi_b(t) - \kappa \beta (t) ,   \\
 \dot c_e (t) & = & -i \frac{g(t)}{\Gamma_r} \beta (t) - i \frac{\Omega^2(t) / \Delta_1}{\Gamma_r} c_e (t) \nonumber \\
 && - i (\Delta_2 - \Delta_1 + \dot \phi_z(t)) c_e (t) ,
\end{eqnarray}
\end{subequations}
where $g(t) = g_c \Omega(t) / \Delta_1$ is the effective coupling rate between $|g\rangle$ and $|e\rangle$ in the absence of decay. These equations resemble Eqs.~(\ref{dotbetat2}) and (\ref{dotce}), which govern the evolution of a two-level system. To see this equivalence, we define new variables $\beta'(t) \equiv \exp[i g_c^2 t/(\Delta_1 \Gamma_r)]\beta(t)$ and $c'_e(t) \equiv \exp[i g_c^2 t/(\Delta_1 \Gamma_r)]c_e(t)$. Assuming that $\gamma_r/\Delta_1 \ll 1$, and thus approximating $|\Gamma_r|^2$ by one, we then obtain 
\begin{subequations}
\label{schadbelm2}
\begin{eqnarray}
\label{schadbelm2_beta}
& \dot \beta' (t) =  - i \frac{g(t)}{\Gamma_r} c'_e (t) - i \sqrt{2 \kappa} \Phi'_b(t) - \kappa \beta' (t) , & \\
\label{schadbelm2_ce}
& \dot c'_e (t) = -i \frac{g(t)}{\Gamma_r} \beta' (t)  + i \Delta c'_e (t) - \gamma c'_e(t) ,&
\end{eqnarray}
\end{subequations}
where
\begin{subequations}
\label{newparam}
\begin{eqnarray}
&\Phi'_b(t)= e^{g_c^2 \gamma_r t/\Delta_1^2} e^{i g_c^2 t /\Delta_1}\Phi_b(t),& \\
& \Delta = [g_c^2-\Omega^2(t)] / \Delta_1 + \Delta_1 - \Delta_2 - \dot \phi_z(t), & \\
& \gamma = g(t)\gamma_r [\Omega(t)/g_c - g_c/\Omega(t)]/\Delta_1 . &
\end{eqnarray}
\end{subequations}
Equations~(\ref{schadbelm2_beta}) and (\ref{schadbelm2_ce}) are similar to Eqs.~(\ref{dotbetat2}) and (\ref{dotce}) if we use the new values for $\gamma$ and $\Delta$ as given by Eq.~(\ref{newparam}), and replace $g$ and $\Phi_b(t)$, respectively, with $g(t)/\Gamma_r$ and $\Phi'_b(t)$. The loading probability is then given by $|c_e(t)|^2 \approx e^{- 2 g_c^2 \gamma_r t/\Delta_1^2} |c'_e(t)|^2$. Note that the term $e^{i g_c^2 t /\Delta_1}$ in $\Phi'_b(t)$ induces a constant frequency shift to the incoming light. We can tilt the center frequency of the input photon from its original value $\omega_0$ to compensate for this induced shift. From now on, we assume that this compensation has been employed. 

For our loading problem, we are looking for particular  pulse shapes for $\Omega(t)$ and  $\phi_z(t)$ that help the $|g\rangle$-to-$|e\rangle$ transition occur with high probability. The traditional solution to this problem employs adiabatic-transfer techniques \cite{stirap, kimble_adiab, nontelep, fleisch}, in which we slowly change the control field so that the system follows its dark-state eigenstate. In our case, for $\Delta_1 = \Delta_2 + \dot \phi_z(t)$ and $\gamma_r = 0$, the system's dark state is given by,
\begin{equation}
|D\rangle = -  \cos \theta(t)  |G\rangle +  \sin \theta(t) |E\rangle .
\end{equation}
where $\cos \theta(t)  \equiv \Omega(t)/\Omega_0$,  $\sin \theta(t)  \equiv g_c/\Omega_0$, and $\Omega_0 \equiv \sqrt{\Omega^2(t)+ g_c^2}$. This state  has the desired property that  for $\Omega(t) \gg g_c$, $|D\rangle \approx - |G \rangle$, but for $\Omega(t) \ll g_c$, $|D\rangle \approx  |E \rangle$. So, if we start with a high value of $\Omega$ when the photon arrives, and then slowly reduce $\Omega$ to zero, we can  adiabatically transfer the system from $|G\rangle$ to $|E\rangle$. The timing is of crucial importance, because if we turn off the pump either before or long after the photon arrives we lose the chance of absorbing the photon. There is also a slight chance of jumping into states that are orthogonal to $|D\rangle$, i.e., $|R\rangle$ or $|B\rangle = \sin \theta(t) |G\rangle + \cos \theta(t) |E\rangle$, in which case, the loading process has completely failed.

Under the dark-state condition, in which the probability of jumping to either $|R\rangle$ or $|B\rangle$ is near zero, we can show that our loading problem is equivalent to loading a cold cavity, viz. a cavity with no atom inside,  with a time-varying decay rate $\kappa \cos^2 \theta(t)$. This can be shown by paralleling the derivation used in \cite{fleisch} and shall be omitted. To maximize the loading probability, we need to find an optimum assignment for $\cos \theta(t)$ that maximizes the loading probability and satisfies our adiabatic conditions. In order to achieve a maximum transfer of free-field photons into the cavity mode, we need to minimize the outgoing field components by destructively interfering the directly reflected and the circulating fields. Fleischhauer et al. have shown that a necessary condition for destructive interference is  \cite{fleisch}:
\begin{equation}
-\frac{d}{dt} \ln \cos \theta (t) + \frac{d}{dt}\ln \Phi_b(t) = \kappa \cos^2 \theta (t).
\end{equation}
For our particular example of a hyperbolic secant  pulse shape $\Phi_b(t) = \sqrt{2/T} {\rm sech} (4t/T)$, it turns out that \cite{fleisch}
\begin{equation}
\label{adbomega}
\Omega(t) =  \frac {g_c {\rm sech} (4t/T)} {\sqrt{[1+ \tanh (4t/T)][\tanh (4t/T)+\kappa T / 2 -1]}}.
\end{equation}
In order for this $\Omega(t)$ expression to yield a  positive  real Rabi frequency, we must have $\kappa T \geq 4$. This, on the other hand, implies that for a successful adiabatic transfer, our input pulse must be long enough so that we can slowly change the quantum memory state.

Here, we propose a simple non-adiabatic approach for loading a $\Lambda$-level atom, which does not impose any restrictions on the input pulse shape and  does not need any adiabatic pulse shaping for the control field.  Our method is based on what we observed in Section~\ref{Sectwolevel} for a two-level atom with a constant coupling rate $g$. There, we realized that a maximum loading probability of greater than 90\% was achievable, provided that we could turn off the atom-light coupling at $t=T_{Load}$. Here, we show that this is indeed possible to do for a $\Lambda$-level atom, by applying a constant control field for an appropriate finite-duration time interval. In other words, we assume $\phi_z(t) =0$ and take $\Omega(t)$ to be
\begin{equation}
\Omega(t) = \left\{ \begin{array}{cc}
\Omega , & \quad\mbox{$t \leq T_{Load}$} \\
0 , & \quad\mbox{$t> T_{Load}$}
\end{array} \right. .
\end{equation}
From Eq.~(\ref{schadbelm2}), the effective coupling rate, $g = g_c \Omega / \Delta_1$, is proportional to the control field's amplitude. By turning off the control field at $t=T_{Load}$,  the coupling rate $g$ vanishes for $t>T_{Load}$. Hence, if  we are in the state $|e\rangle$ at $t=T_{Load}$, we will stay there until a decay process returns the atom to its ground state. 
Solving for $c'_e(t)$ in Eq.~(\ref{schadbelm2_ce}) using Eq.~(\ref{cet}),
the loading probability is then given by
\begin{eqnarray}
\label{cetlmd}
\lefteqn{|c_e(t)|^2 \approx \frac{2 \kappa g^2  } { |\xi|^2 } \times} \nonumber \\
&& \left|  \int_0^t { d \tau  e^{-g_c^2 \gamma_r (t-\tau)/ \Delta_1^2} \Phi_b (\tau) (e^{-\kappa_+ (t- \tau)} - e^{ - \kappa_- (t- \tau)})} \right|^2 , \nonumber \\
&& \hspace{1in} \quad\mbox{for $t \leq T_{Load}$,}
\end{eqnarray}
where $\kappa_\pm$ and $\xi$ are given by Eq.~(\ref{loadingparam}) using   $\gamma$ and $\Delta$ as given by Eq.~(\ref{newparam}). The loading probability is then given by $|c_e(T_{Load})|^2$. The effective detuning, $\Delta = (g_c^2 - \Omega^2)/ \Delta_1 + \Delta_1 - \Delta_2$,  can be forced to zero by properly choosing $\Delta_1$ and $\Delta_2$, thus compensating for the induced Stark shift and enabling us to obtain a higher loading probability.

\subsubsection{Adiabatic versus non-adiabatic loading: numerical comparison}

In this section, we compare our proposed non-adiabatic scheme to loading mechanisms that use adiabatic-transfer techniques. Adiabatic transfer is a well-studied problem in the literature. To make our comparison explicit,  we will consider the particular examples of adiabatic loading that have been proposed in \cite{nontelep} and \cite{fleisch}. In \cite{nontelep}, the authors have devised a method for transferring the state of a trapped-atom quantum memory to another trapped-atom quantum memory. Their approach is based on adiabatically transferring the state of one memory to a single photon, which then propagates to and loads the other memory. Their loading process is facilitated by forcing the photon's pulse shape to be symmetric, so that the receiver can employ a time-reversed version of the control pulse that was used at the transmitter. The desired pulse shape for the control field, under the dark-state condition at zero effective detuning, can be found numerically by solving the corresponding Schr\"odinger equations. However, the approach from \cite{nontelep} is not suitable  for an incoming photon with an arbitrary pulse shape. In \cite{fleisch}, the authors have employed the adiabatic transfer  to load an atomic ensemble with the state of a single photon. They have  employed the dark-state approximation under the two-photon resonance condition with a constant-phase control field. Their approach sets certain limitations on the length of the input pulse shape. In fact, in order to fulfill the dark-state condition, the input pulse shape must be longer than a threshold value. Unlike these two adiabatic mechanisms, our non-adiabatic approach puts no constraints on the input pulse shape.

Several issues make our work in this paper, and the numerical results presented in this section,  distinct from previous work on loading quantum states into neutral atoms. The first issue is our accounting for a nonzero probability for populating the bright state $|B\rangle$. We will see in this section how the nonzero probability of being in the bright state affects the loading performance. That also makes it possible to scrutinize the dependence of the loading probability, in both adiabatic and non-adiabatic mechanisms, on the key system parameters, e.g., $g_c/\kappa$ and $\kappa T$. This evaluation is one of the original contributions of this paper. Finally, our use of a constant control field allows us to obtain analytical results for loading dynamics in the non-adiabatic case.

It is of practical importance to know at what values of input bandwidth, represented by $\kappa T$, and atomic coupling rate, represented by $g_c/\kappa$, a desired system performance can be achieved.
For our non-adiabatic approach, at $\gamma_r = \Delta = 0$, system performance is governed by the  effective coupling rate $g = g_c \Omega / \Delta_1$. The optimum value of $g/\kappa$ can then be obtained by making appropriate choices for $g_c$, $\Omega$, and $\Delta_1$. The only conditions that we need to satisfy are $\Delta_1 \gg \gamma_r$ and $g_c,\Omega \ll \Delta_1$. That allows us some flexibility to pick a smaller  $g_c$, which determines the cavity length, and a larger $\Omega$, than is  the case for the adiabatic approach proposed in \cite{fleisch}. For instance, using the hyperbolic secant  pulse shape and the optimum control pulse shape, given by Eq.~(\ref{adbomega}), we have $g_c \Omega(t) = g_c^2 \Omega'(t)$, where $\Omega'(t)$ is only a function of $\kappa T$ and not $g_c$. Therefore, for a fixed value of $\kappa T$, the only way to increase the coupling rate in the adiabatic case is to use a shorter cavity, which  yields a larger $g_c$.

For nonzero values of $\gamma_r/\Delta_1$, there is another parameter that we can vary  to optimize the loading probability in our non-adiabatic scheme: it is $g_c/\Omega$, which appears  in  the exponents $ -[g_c^2 \gamma_r/\Delta_1^2 + \kappa_\pm] (t-\tau)  = -[ \kappa + g (g_c/\Omega+\Omega/g_c)(\gamma_r/\Delta_1) \pm \xi](t-\tau)/2$ in Eq.~(\ref{cetlmd}). To increase the loading probability, for  fixed values of $g/\kappa$ and $\gamma_r/\Delta_1$, we need low values of  these exponents. These can be achieved at $g_c/\Omega \approx 1$. The exact optimum value for $g_c / \Omega$  that minimizes the decay effect also depends on the values of $\gamma$ and $\xi$. For the sake of comparison, however, we assume that $\gamma_r/\Delta_1$ is sufficiently small that we can neglect these subtleties, and we only focus on the interplay between the coupling rate and the input bandwidth. So, for the rest of this section, we assume $\gamma_r = 0$.

\begin{figure}
	\centering
		\includegraphics  [width = 0.8\linewidth] {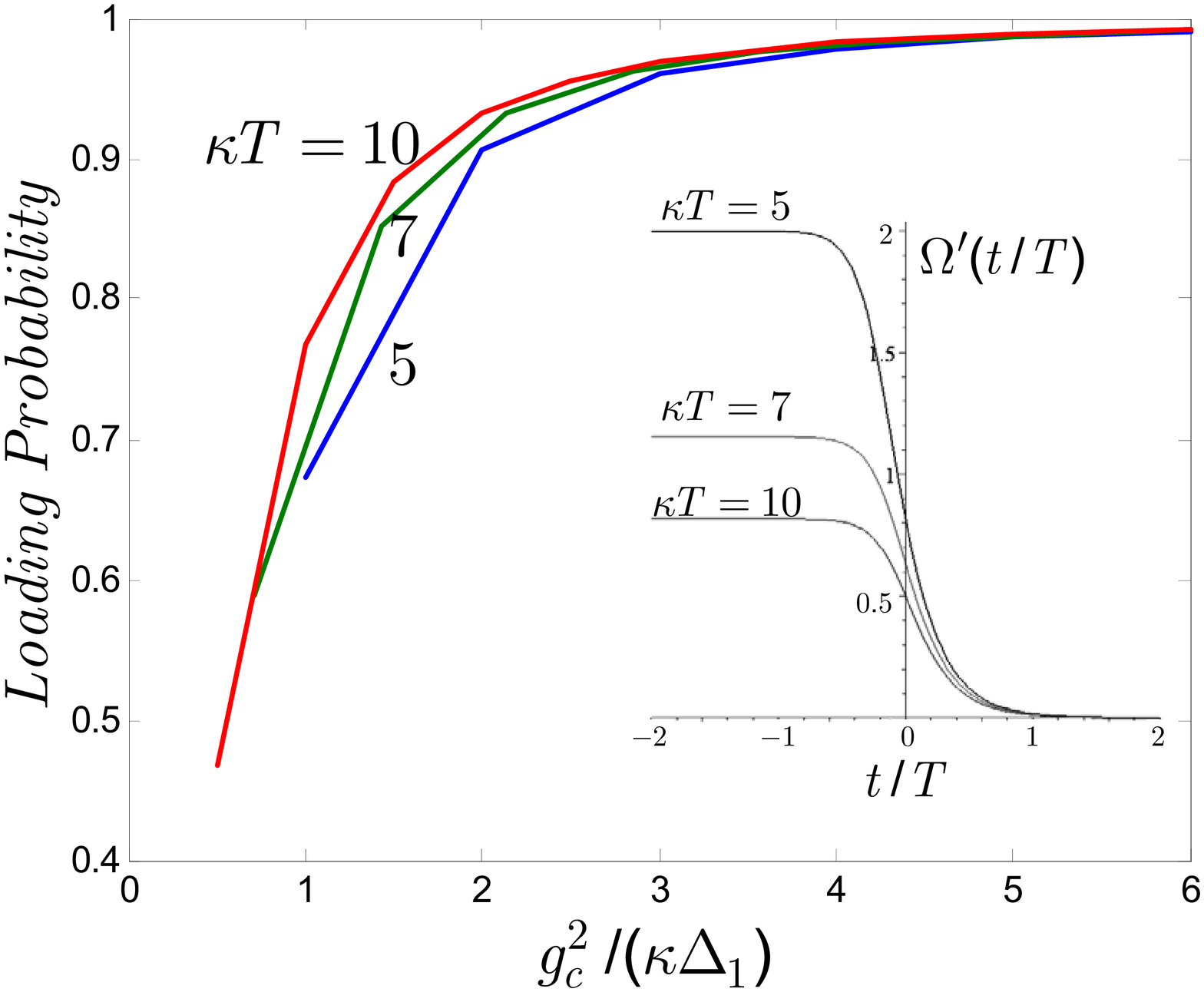}
	\caption{(Color online) The loading probability for the adiabatic method as a function of $\kappa T$ versus $g_c^2/(\kappa \Delta_1) $, at the two-photon resonance $\Delta_1 = \Delta_2$ with $\phi_z(t) = 0$, $\gamma_r = 0$, and several values of $\kappa T$. These curves were obtained by numerically solving Eq.~(\ref{schadbelm}) using the control pulse shapes, shown in the inset, that correspond to a hyperbolic secant  input pulse shape.}
	\label{adbtwores} 
\end{figure}

It is interesting to find the dependence of the loading probability, for the adiabatic scheme proposed in \cite{fleisch}, on the coupling rate $g_c$. For this purpose, we have numerically solved the Schr\"odinger equations in Eq.~(\ref{schadbelm}), for $\Omega(t)$ given by Eq.~(\ref{adbomega}), at the two-photon resonance $\Delta_1 = \Delta_2$  and for $\phi_z(t) = 0$. In Fig.~\ref{adbtwores} we have plotted $|c_e(5T)|^2$ versus an effective coupling rate $g' \equiv g_c^2/ \Delta_1$. Choosing $t=5T$ ensures that the loading process has ended, and therefore $|c_e(5T)|^2$ is effectively the loading probability. This figure shows that for the two-photon resonance case, higher $g_c$ values yield higher loading probabilities. The effective coupling rate $g'$ that we need is about $2 \kappa$ for 90\% loading probability at $\kappa T =5$. It can be seen that there is an advantage to using pulses with higher values of $\kappa T$,  because our adiabatic scheme is more efficient for longer input pulses.

\begin{figure}
	\centering
		\includegraphics  [width = .7\linewidth] {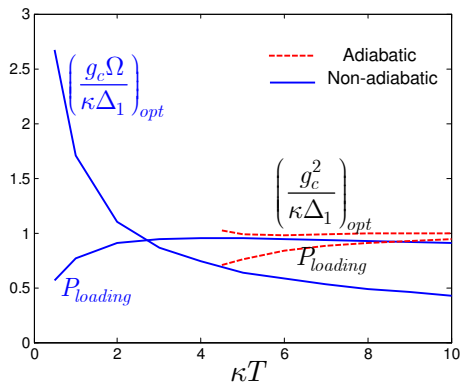}
	\caption{(Color online) The optimum coupling rate and the maximum loading probability achieved at this rate for both adiabatic and non-adiabatic approaches. (Non-adiabatic curves start from $\kappa T = 0.5$; adiabatic curves start from $\kappa T =4.5$.) For all curves, we assume that $\Delta = \gamma_r = 0$. A hyperbolic secant  pulse shape has been used for the incoming photon.}
	\label{designp2} 
\end{figure}

We can also solve the Schr\"odinger equations in Eq.~(\ref{schadbelm}), for $\Omega(t)$ given by Eq.~(\ref{adbomega}), at zero effective detuning, i.e., when $\dot \phi_z(t) = \Delta_1 - \Delta_2 + [g_c^2 - \Omega^2(t)]/ \Delta_1$, as proposed in \cite{nontelep}. Note that the control field's Rabi frequency given by Eq.~(\ref{adbomega}) has been obtained under the two-photon resonance and dark-state conditions, and it is therefore not  the optimum pulse shape for the zero-effective-detuning case. It can be verified, numerically, that this control field provides a close-to-optimum performance. This is in accord with the results reported in \cite{fleisch} implying that the loading probability will only be  slightly affected by small deviations from the optimum control field pulse shape. For zero-effective detuning, in analogy to the non-adiabatic case, we observe the existence of an optimum value for $g'/\kappa$. Figure~\ref{designp2} shows the optimum coupling rates and the maximum loading probabilities, $P_{loading}$, achieved at these rates as functions of $\kappa T$ for the adiabatic and non-adiabatic approaches using a hyperbolic secant  pulse shape. Here, $(g'/\kappa)_{opt}$ is a near-unity constant, for all values of $\kappa T$, whereas $(g/\kappa)_{opt}$ decreases toward zero with increasing $\kappa T$. The maximum loading probability that can be achieved by the adiabatic approach asymptotically goes to one as $\kappa T$ increases. Its value is about 75\%  for $\kappa T = 4.5$, but we should bear in mind that the system will be more vulnerable to spontaneous decay when we use long pulses. 

Because the non-adiabatic approach has no constraint on input pulse width, it allows a larger class of single-photon sources to be used in quantum communication systems, which can result in  higher throughputs. For instance, consider system $A$, which uses a downconverter single-photon  source  at $\kappa T = 1$, and  non-adiabatic loading  at the optimum coupling rate. We will compare it with  system $B$, which uses adiabatic loading with the same source but at $\kappa T =4$, the minimum required for the adiabatic case in our examples. Both  systems use identical quantum memory modules, so their $\kappa$ values are the same. The total throughput is proportional to the source rate and the loading probability. The source rate for system $A$ is four times that of system $B$, because we are allowing a larger output bandwidth in the former case. From Fig.~\ref{designp2}, the maximum achievable loading probability for system $A$ is above 75\%, while that for system $B$ is at most one. It folllows that system $A$ can have a $4 \times 0.75 = 3$ times higher througput than system $B$.

\begin{figure}
	\centering
		\includegraphics  [width = .745\linewidth] {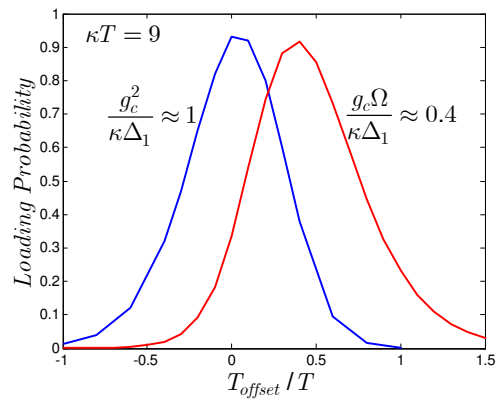}
	\caption{(Color online) Loading probability versus normalized timing offset in the photon arrival. In the adiabatic case (the left curve), this timing error results in applying the control field earlier or later than the correct time. In the non-adiabatic case (the right curve), it results in stopping the control field earlier or later than $T_{Load}$. Both curves assume $\Delta = \gamma_r = 0$ for their respective optimum coupling rates for hyperbolic secant  input pulse shapes.}
	\label{timesens} 
\end{figure}

Finally, Fig.~\ref{timesens} shows the sensitivity of both schemes to timing offsets. Here, $T_{\rm offset}$ refers to the time that we stop the control field in the non-adiabatic approach, and it represents a time shift in the adiabatic approach, i.e., using $\Omega(t-T_{\rm offset})$ instead of $\Omega(t)$. Both cases may occur if we have an inaccurate estimate of the photon arrival time. This figure shows that both schemes have almost the same tolerance for timing offsets. A 50\% loading probablity (3~dB loss) is achievable even if we are about $\pm T/2$ off from the correct loading time. Although it is not shown in this figure, the same result holds if we use the adiabatic approach under the two-photon-resonance condition.

\subsection{Double-$\Lambda$ Trapped Atoms}
\label{SecVlevel}

\begin{figure}
	\centering
		\includegraphics  [width = \linewidth] {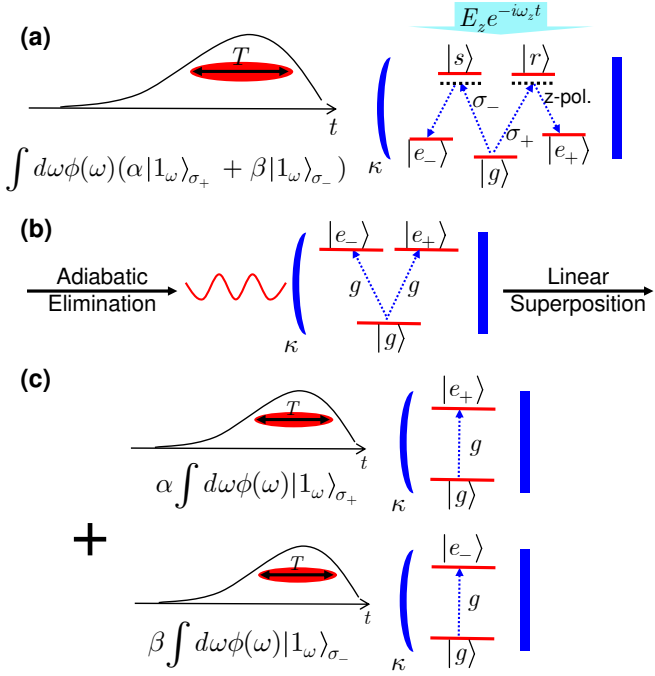}
	\caption{(Color online) (a) Loading a double-$\Lambda$ atom driven by a polarization-state single photon. By adiabatically eliminating the upper states, this system becomes equivalent to the $V$-level system shown in (b). (c)  Reduction of the $V$-level loading problem to two two-level loading problems by means of linear superposition.}
	\label{dbllam} 
\end{figure}

The results we obtained above are directly applicable to several other problems. For instance, consider the double-$\Lambda$ atom shown in Fig.~\ref{dbllam}(a). Here, we are trying to load the memory with a single photon in an arbitrary polarization state:
\begin{equation}
\label{dblinit}
|\psi_0\rangle = \int {d \omega \phi(\omega) (\alpha |1_\omega\rangle_{\sigma_+} + \beta |1_\omega\rangle_{\sigma_-}) },
\end{equation}
where $|\alpha|^2 + |\beta|^2 =1$, and $|1_\omega\rangle_{\sigma_\pm}$ refers to a single-photon state at frequency $\omega$  in the $\sigma_\pm$ (right/left circular) polarization. The goal of the loading process is to absorb this photon, by properly choosing $z$-polarized control fields---according to either the adiabatic or non-adiabatic  mechanisms---and store its polarization information in the metastable levels $|e_+\rangle$ and $|e_-\rangle$  that correspond  to the $\sigma_+$ and $\sigma_-$ polarizations, respectively. The loading probability is therefore determined by the probability of ending up in $|e\rangle \equiv \alpha |e_+\rangle + \beta |e_-\rangle$. 

In order to calculate the loading probability, we can use the results of the previous subsection to reduce each $\Lambda$-configuration leg of the atom to an equivalent two-level system, as shown in Fig.~\ref{dbllam}(b). The resulting configuration is a $V$-level atom, illuminated by an arbitrary-polarized single photon. This problem can  be solved by noting that the full Schr\"odinger equations---for the state of the reservoirs, cavity modes, and the atom---that model the system evolution are  linear in response to the superposition state in Eq.~(\ref{dblinit}). Therefore, each leg of the $V$-level atom is only driven by the corresponding part in the initial state, as shown in Fig.~\ref{dbllam}(c). In fact, for $\alpha \int{d \omega \phi(\omega) |1_\omega\rangle_{\sigma_+}}$ as the input, $c_-(t)$---the  probability amplitude of being in the state $|e_-\rangle$---will be zero. We only get a nonzero $c_-(t)$ from the input $\beta \int{d \omega \phi(\omega) |1_\omega\rangle_{\sigma_-}}$, in which  case the $c_-(t)$ evolution coincides with that of a two-level system. For instance, if we are using the non-adiabatic loading mechanism, $c_-(t)$ is given by $\beta c_e(t)$, using $c_e(t)$ from Eq.~(\ref{cetlmd}). The same argument holds for $c_+(t)$, so that the overall probability of ending up in $|e\rangle$ is  $(|\alpha|^2 + |\beta|^2)|c_e(t)|^2 = |c_e(t)|^2$. This proves the equivalence of loading a $V$-level atom to the loading of a two-level atom when both are driven by single photons. This result  also extends to cases in which the two legs of the atomic system are nonlocal---e.g., when a pair of two-level trapped-atom memories are driven by a photon-number entangled state. Such a system is an example of the to-the-memory configuration for entanglement distribution, and it can be used in teleportation systems. 

\section{MIT-NU Entanglement Distribution} 
\label{Secentangle}

With the results that we have developed for trapped-atom quantum memories in the previous section, we can address the performance of quantum communication systems that use this type of storage. Our focus on this section will be on the MIT-NU architecture for quantum communication  \cite{MIT/NU}. The MIT-NU construct uses the to-the-memory configuration in Fig.~\ref{telepapp}(a) to distribute and share polarization entanglement between two rubidium atoms. Because the rubidium atoms can be treated as double-$\Lambda$ quantum memories \cite{trap},   MIT-NU entanglement distribution is  essentially a loading problem, in which we are interested in transferring an entangled state to two trapped atoms. 
All previous analyses of the MIT-NU architecture \cite{ACM,MIT/NU,Brent}, however, have employed a cold-cavity approach in which each optical cavity---that would hold an ${}^{87}$Rb atom in the actual implementation---is regarded as empty, and the loading probability is calculated by determining the probability that  the state of the intracavity photon fields at the end of a loading interval is the desired singlet. In this section, we provide the first loading analysis of the MIT-NU system that includes the presence of atoms within the quantum memory units.

\begin{figure}
   \centering
      \includegraphics[width = \linewidth]{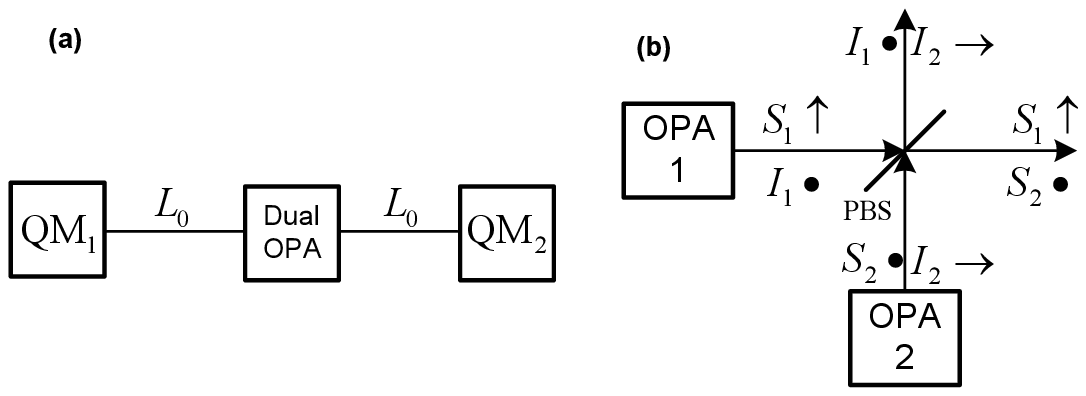}
      \includegraphics[width = 0.5\linewidth]{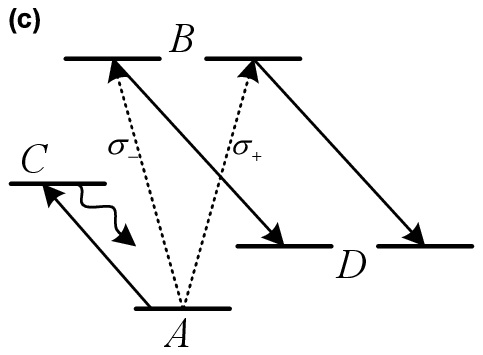}
      \caption
   { \label{FM1} 
(a) MIT-NU architecture for long-distance quantum communications consisting of a dual-OPA source that produces polarization-entangled photons, and two quantum memories, QM$_1$ and QM$_2$, separated by $2L_0$\,km. (b) Dual-OPA source of polarization-entangled photons.  OPAs 1 and 2 are coherently-pumped ($\pi$-rad out of phase), continuous-wave, type-II phase matched, doubly-resonant amplifiers operated at frequency degeneracy whose orthogonally-polarized signal ($\{S_k\}$) and idler ($\{I_k\})$ outputs are combined, as shown, on the polarizing beam splitter (PBS). (c) Notional schematic for the relevant hyperfine levels of ${}^{87}$Rb. Each quantum memory consists of a single trapped rubidium atom that can absorb arbitrarily polarized photons, storing their coherence in the long-lived $D$ levels. A non-destructive load verification is effected by means of the $A$-to-$C$ cycling transition.}
   \end{figure}

Figure~\ref{FM1}(a) shows a schematic of the MIT-NU system:  QM$_1 $ and QM$_2$ are trapped rubidium atom quantum memories, each $L_0 $\,km away---in opposite directions---from a dual optical parametric amplifier (OPA) source. Each optical parametric amplifier in the dual-OPA source is a continuous-wave, type-II phase matched, doubly-resonant amplifier operating at frequency degeneracy $\omega_S=\omega_I$.  Its signal ($S$) and idler ($I$) outputs comprise a stream of orthogonally-polarized photon pairs that are in a joint Gaussian state \cite{MIT/NU}.  By coherently pumping two of these OPAs, $\pi$-rad out of phase, and combining their outputs on a polarizing beam splitter, as shown in Fig.~\ref{FM1}(b), we obtain signal and idler beams that are polarization entangled \cite{ultrabright}. Moreover, double-resonant operation greatly enhances the brightness of these beams, making this entanglement source compatible with loading atomic memories \cite{chris}, something that is not the case for standard $\sim$THz bandwidth parametric down-converter sources.  The signal and idler beams are routed down separate optical fibers to the trapped-atom quantum memories.  To do so efficiently, their wavelength is chosen to lie in the low-loss 1.55\,$\mu$m band.  However, because the quantum memory makes use of the $^{87}$Rb line at 795\,nm, quantum-state frequency conversion is used, at each memory location, to convert the polarization entanglement from 1.55\,$\mu$m to 795\,nm \cite{Albota}.  As explained in \cite{Brent}, additional steps are taken to, prior to the upconversion, to compensate for any polarization transformation encountered in transmission from the source to the memory.  This is accomplished by monitoring the polarization change incurred on a strong pilot pulse and using that information to drive polarization controllers that compensate the signal and idler beams.  

A schematic of the relevant hyperfine levels of ${}^{87}$Rb is shown in Fig.~\ref{FM1}(c).  The memory atoms are initially in the ground state $A$.  From this state they can absorb a photon in an arbitrary polarization  and, by means of a Raman transition, transfer that photon's coherence to the long-lived $D$ levels  for subsequent use.  However, because propagation and fixed losses may destroy photons before they can be stored, and because both memories must be loaded with a singlet state prior to performing qubit teleportation, the MIT-NU architecture employs a clocked loading protocol in which the absence of fluorescence on the $A$-to-$C$ cycling transition provides a non-destructive indication that a memory atom has absorbed a photon.  If no fluorescence is seen from either the QM$_1$ or QM$_2$ atoms in a particular loading interval, then we assume that both memories  have stored photon coherences and so are ready for the rest of the teleportation protocol, i.e., Bell-state measurement, classical communication of the result, and single-qubit rotation \cite{trap}. No entanglement purification step was envisaged in the original description of the MIT/NU architecture, although one certainly could be added if multiple pairs are stored at each location.

In order to find analytical results for the MIT-NU loading problem, we make several simplifying assumptions. First, we replace the OPA devices in the entanglement source with simple spontaneous parametric downconverters (SPDCs). This is actually how most realizations of this system work \cite{PDC3,PDC4}, and,  except for their flux and their output bandwidths, SPDC systems have the same physical characteristics as OPA sources. The second assumption is the {\em biphoton} approximation to the SPDC output. The latter is mostly a vacuum state, plus a small biphoton component, and a much smaller  multi-pair contribution \cite{MIT/NU}. The vacuum term can be easily recognized by our non-destructive loading verification technique, and can therefore be ignored. The multi-pair case is  of minor concern in the context of loading, because its degrading effect has been accounted for in the previous analyses \cite{Brent}. So, in this section, we only consider the case in which the type-II phase-matched SPDC output, operating at frequency degeneracy, is a biphoton state in the following general form \cite{TBDB}
\begin{equation}
|\psi\rangle = \int {d \omega_S \int {d \omega_I \phi(\omega_S,\omega_I) |1_{\omega_S}\rangle_{S,\diamond} |1_{\omega_I}\rangle_{I,\bar \diamond} }},
\end{equation}
where $\int {d \omega_S \int {d \omega_I |\phi(\omega_S,\omega_I)|^2 }} = 1$, and $\diamond$ and $\bar \diamond$ represent two orthogonal polarizations. For the particular configuration in Fig.~\ref{FM1}(b), the desired output state will then be
\begin{equation}
|\psi_{out}\rangle  =  \int {d \omega_S \int {d \omega_I \phi(\omega_S,\omega_I) \left| \psi^-_{\omega_S,\omega_I} \right \rangle_{SI} }} ,
\end{equation}
where
\begin{equation}
\left| \psi^-_{\omega_S,\omega_I} \right \rangle_{SI} \equiv
\frac{ |1_{\omega_S}\rangle_{S,\sigma_+} |1_{\omega_I}\rangle_{I,\sigma_-} -|1_{\omega_S}\rangle_{S,\sigma_-} |1_{\omega_I}\rangle_{I,\sigma_+}  }{\sqrt{2}}
\end{equation}
is   the singlet state, which is invariant with respect to polarization basis. Finally, we assume that there is no loss and no delay in the channel so that  the preceding state is the input state to the quantum memories.

   \begin{figure}
   \centering
      \includegraphics[width=\linewidth]{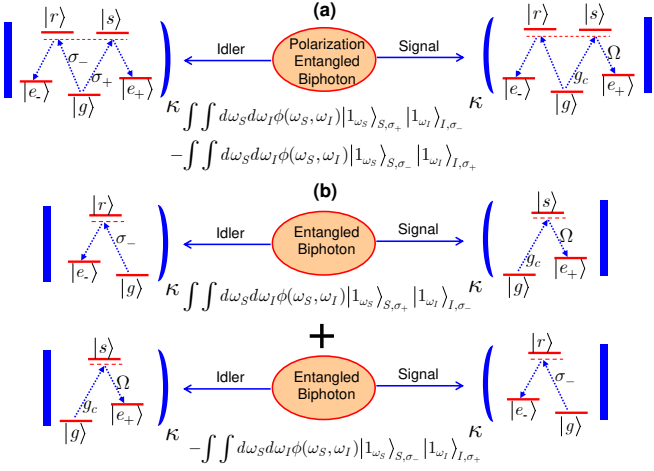}
      \caption
   { \label{MITNUld} 
(Color online) (a) Loading a pair of double-$\Lambda$ atoms illuminated by a polarization-entangled biphoton. This is a fair approximation to the MIT-NU loading problem. (b) Breaking the loading problem in  (a) into two simpler loading problems for a pair of $\Lambda$-level atoms,  each illuminated by a biphoton state.}
   \end{figure} 

Figure~\ref{MITNUld}(a)  shows a schematic of the MIT-NU loading problem with the above considerations taken into account. Here, similar to the previous section, we assume that the signal and idler photons are spatially matched to one of the cavity's spatial modes with center frequency $\omega_0$.  Depending on their polarizations, they can drive one of the transitions $|g\rangle \rightarrow |s\rangle$ (for $\sigma_+$-polarization) and $|g\rangle \rightarrow |r\rangle$ (for $\sigma_-$-polarization) with time-independent coupling rate $g_c$ and detuning $\Delta_1$. The second transition is facilitated by a $z$-polarized classical field, which induces a Rabi frequency $\Omega$ with detuning $\Delta_2$. The goal of memory loading is to absorb the entangled photons and store their coherence in the metastable levels of each memory, i.e., to end up in the state $|e_+\rangle_S |e_-\rangle_I - |e_-\rangle_S |e_+\rangle_I$, where the subscript $S/I$ refers to the  atom that is driven by the signal/idler photon. 

The above problem can be reduced to the two simpler problems, shown in Fig.~\ref{MITNUld}(b), by using the linearity in the input superposition state. This is the same technique that we used in Section~\ref{SecVlevel} to reduce a $V$-level atom to a pair of two-level atoms. Now that our problem has been reduced to the loading problem for two $\Lambda$-level atoms, we can employ the adiabatic and non-adiabatic loading mechanisms described in Section~\ref{Seclambda}. We provide analytical results for the loading probability associated with either of the systems shown in Fig.~\ref{MITNUld}(b), which can be directly applied to the case shown in Fig.~\ref{MITNUld}(a), or equivalently, to our approximation to the MIT-NU loading problem. Because the two subsystems in Fig.~\ref{MITNUld}(b) have the same loading behavior---independent of the polarization of the incoming photons---we simplify our notation  by omitting the polarization information. The  driving state will then be written as follows
\begin{equation}
\label{ini_state_res}
| \psi _ 0 \rangle = \int { {d}  \omega_S \int { { d}  \omega_I \phi(\omega_S, \omega_I ) |1_{\omega_S}  \rangle_S |1_{\omega_I} \rangle_I}},
\end{equation}
where $\int { {d}  \omega_S \int { { d}  \omega_I |\phi(\omega_S, \omega_I )|^2}} = 1$,  and the subscripts $S$ and $I$ refer to the two independent reservoirs that interact with our memory cells. 

\begin{figure}
   \centering
      \includegraphics[width=\linewidth]{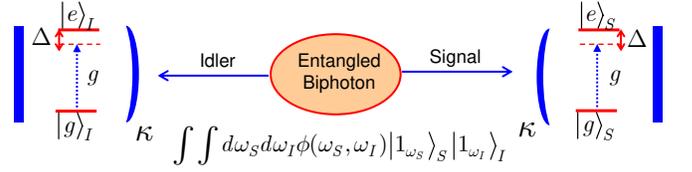}
      \caption
   { \label{MITNUnonadbfig} 
(Color online) A two-level-atom model for the systems shown in Fig.~\ref{MITNUld}(b).  Here, we have assumed that the Rabi frequency $\Omega$ is a constant  and we have adiabatically eliminated the upper state in the $\Lambda$-level atoms. In the absence of spontaneous decay, the effective coupling rate is $g= g_c \Omega /\Delta_1$. The effective detuning $\Delta$ can be made zero by a proper choice of parameters, as shown in Eq.~(\ref{schadbelm}).}
   \end{figure}

In this section, we only derive the loading probability for our non-adiabatic scheme. 
As we showed in Eq.~(\ref{schadbelm2}), if the decay rates from the upper states are much smaller than the employed detuning, each $\Lambda$-level atom can be approximated by a two-level atom with an effective coupling rate $g=g_c \Omega / \Delta_1$ and a detuning $\Delta$ that can be made zero by a proper choice of parameters. This reduces the loading problem in Fig.~\ref{MITNUld}(b) to the one shown in Fig.~\ref{MITNUnonadbfig}, in which a pair of trapped two-level atoms are illuminated by a biphoton state $|\psi_0\rangle$ as given by Eq.~(\ref{ini_state_res}). Here, the loading event corresponds to  populating both atomic excited states, i.e., to ending up in the state $|e\rangle_S |e\rangle_I$.

To analyze this loading problem,  we can again use the superposition trick from the previous subsection to find $c_{ee}(t)$, the slowly-varying probability amplitude at time $t$ associated with $|e\rangle_S |e\rangle_I$. Using $c_{\omega_S,\omega_I}(t)$ to denote the slowly-varying probability amplitude at time $t$ associated with $|e\rangle_S |e\rangle_I$ in response to the two-photon state $\phi(\omega_S, \omega_I ) |1_{\omega_S}  \rangle_S |1_{\omega_I} \rangle_I$, we have
\begin{equation}
c_{ee}(t)  = \int { {d}  \omega_S \int { { d}  \omega_I c_{\omega_S,\omega_I}(t) }},
\end{equation}
where, from Eq.~(\ref{ceomegat}),
\begin{eqnarray}
c_{\omega_S,\omega_I}(t) {d}  \omega_S {d}  \omega_I &=& \frac{\kappa g^2}{\pi \xi^2}  \int_0^t { d \tau \int_0^t { d \tau' d \omega_S d \omega_I \phi(\omega_S,\omega_I) }}  \nonumber \\
&& \times e^{-i(\omega_S-\omega_0)\tau}  
 e^{-i(\omega_I-\omega_0)\tau'} \nonumber \\
&& \times (e^{-\kappa_+ (t- \tau)} - e^{ - \kappa_- (t- \tau)}) \nonumber \\
&& \times (e^{-\kappa_+ (t- \tau')} - e^{ - \kappa_- (t- \tau')}) .
\end{eqnarray}
The above equation results in the following expression for the time-dependent probability amplitude  for being in the state $|e\rangle_S |e\rangle_I$:
\begin{eqnarray}
\label{betaeet}
 c_{ee}(t)  & = & \frac{\kappa g^2}{\pi \xi^2}  \int_0^t { d \tau \int_0^t { d \tau' \Phi_b(\tau,\tau') }} \nonumber \\
 && \times (e^{-\kappa_+ (t- \tau)} - e^{ - \kappa_- (t- \tau)})  \nonumber \\
&&  \times (e^{-\kappa_+ (t- \tau')} - e^{ - \kappa_- (t- \tau')}) ,
\end{eqnarray}
where
\begin{equation}
\Phi_b(\tau, \tau') = \int {\hspace{-.07in} d \omega_S \int  {\hspace{-.07in}  d \omega_I e^{ - i  (\omega_S - \omega_0) \tau - i  (\omega_I - \omega_0) \tau'}\phi( \omega_S,   \omega_I) }}.
\end{equation}
 Note that in the above integral only the symmetric part of the pulse shape, $[\Phi_b(\tau,\tau') + \Phi_b(\tau',\tau)]/2$, results in a nonzero value for $c_{ee}(t)$. Also, it has been implicitly assumed that the pulse shape $\Phi_b(\tau,\tau')$ is nonzero only for $\tau, \tau' >0$. Otherwise we have to change the lower limits of the above double integral.

\subsection{Numerical results}

Here, we present some numerical results for the performance of MIT-NU non-adiabatic loading mechanism. Our goal  is to find the dependence of the loading probability on the bandwidth of the input pulse as well as on the atom-light coupling rate.  This loading probability is given by $|c_{ee}(t)|^2$ as obtained in Eq.~(\ref{betaeet}). The input pulse shape that we consider here corresponds to the output of an SPDC, which is given by \cite{TBDB}
\begin{equation}
\label{phispectral}
\phi(\omega_S,\omega_I) = A \phi_+(\omega_S,\omega_I) \phi_-(\omega_S,\omega_I).
\end{equation}
In this expression, $A$ is a normalization factor, and
\begin{eqnarray}
\phi_+(\omega_S,\omega_I) & = & \frac{\sqrt{\omega_S \omega_I}}{n_S(\omega_S) n_I(\omega_I)} \phi_P(\omega_S+ \omega_I) \nonumber \\
& \cong & \frac{\omega_0}{n_S(\omega_0) n_I(\omega_0)} \phi_P(\omega_S+ \omega_I),
\end{eqnarray}
where $n_{S/I}$ is the downconversion crystal's refractive index for the signal/idler beam and $\phi_P(\omega)$ is the pump spectral pulse shape. The other term in Eq.~(\ref{phispectral}) is the phase-matching function,
\begin{equation}
\phi_-(\omega_S , \omega_I) = \frac {\sin [\Delta k (\omega_S , \omega_I) \, L/2]} {\Delta k (\omega_S , \omega_I) /2},
\end{equation}
where $L$ is the crystal length, and $\Delta k (\omega_S , \omega_I) \equiv k_P(\omega_S+ \omega_I)- k_S(\omega_S) - k_I(\omega_I)$, with $k_{S/I/P}$ being the  wave number of the signal/idler/pump beam. We assume that the crystal is phase matched at degeneracy, i.e., $k_P(2 \omega_0) = k_S(\omega_0) + k_I(\omega_0)$, and that it also satisfies  the extended phase-matching condition, $ k'_P(2 \omega_0) - k'_S(\omega_0) = k'_I (\omega_0) - k'_P(2 \omega_0) \equiv \delta k $, where $k'_X(\omega)$, $X=S,I,P$, is  the derivative of $k_X$ with respect to $\omega$. A first-order linear approximation will then yield $\Delta k (\omega_S , \omega_I) \approx (\omega_S - \omega_I) \delta k$.
Consequently, $\phi_\pm(\omega_S, \omega_I)$ is only a function of  $\omega_S \pm \omega_I$.  The temporal baseband  pulse shape needed for our loading probability calculation is therefore
\begin{equation}
\Phi_b(t,u)   =  \frac{A'}{2}  \Phi_{b} \left( \frac{t+u}{2} \right) \Phi_{-} \left( \frac{t-u}{2} \right) ,
\end{equation}
where $A' = A \omega_0/[n_S(\omega_0) n_I(\omega_0)]$, $\Phi_b(t) \equiv e^{2 i \omega_0 t} \int {d \omega \phi_P(\omega)e^{-i \omega t}}$ is the baseband pump pulse shape, and $\Phi_-(t)  = (2\pi / \delta k) [u(t+T_0/2)-u(t-T_0/2)]$, where $u(t)$ is the step function and $T_0 = \delta k \, L$.

\begin{figure}
\centering
\begin{tabular} {c}
\includegraphics[width=.7\linewidth]{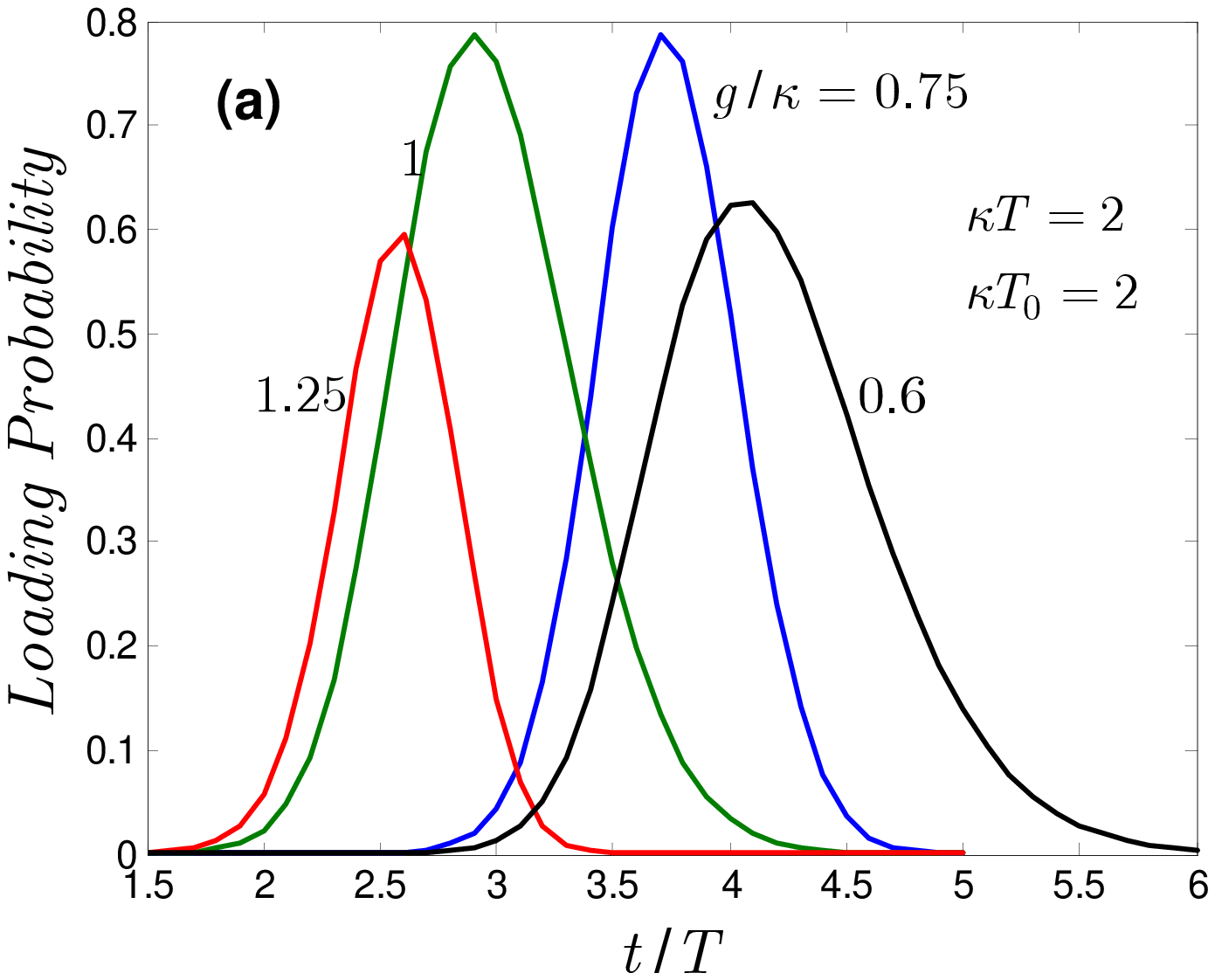} \\
\includegraphics[width=.7\linewidth]{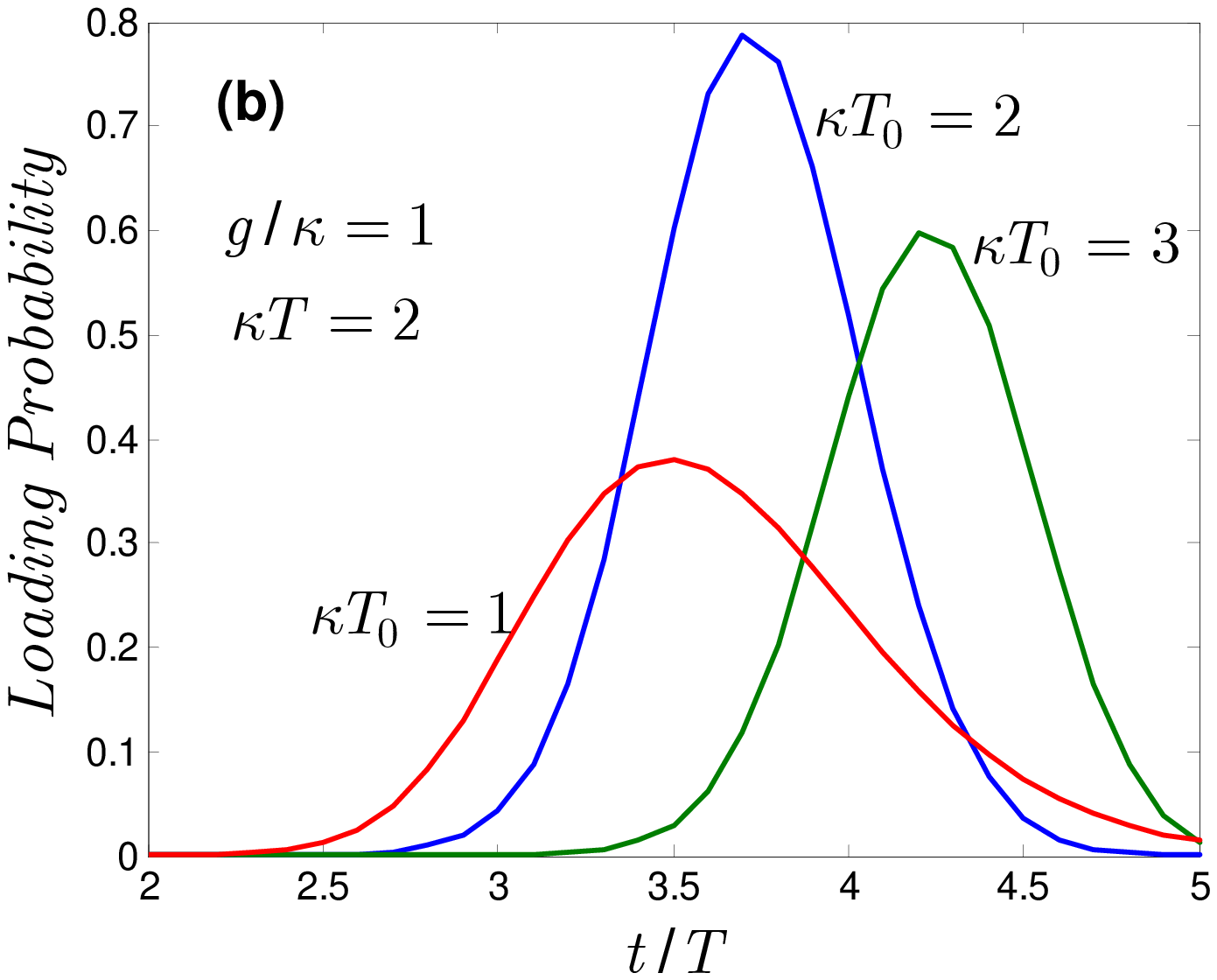} 
\end{tabular}
 \caption{ \label{MITNUloadpr} (Color online)
(a) MIT-NU loading probability for different values of $g/\kappa$ at $\kappa T = \kappa T_0 =2$. The maximum probability in each case is a function of $g/\kappa$, which implies the existence of an optimum value for $g/\kappa$. (b) MIT-NU loading probability for different values of $\kappa T_0$ at $g/\kappa =1$ and $\kappa T = 2$. In all curves, we have used a hyperbolic secant  pulse shape for the pump beam with $\gamma'=0$.}
\end{figure}

Figure~\ref{MITNUloadpr} shows the loading probability for the architecture in Fig.~\ref{MITNUnonadbfig} when hyperbolic secant  pulse shape $\Phi_b(t) = \sqrt{2/T} {\rm sech} [4(t- T_0-2 T)/T]$ is employed. We have shifted the pump pulse shape by $2T + T_0$ so that, effectively, $\Phi_b(t,u) = 0$ for $t<0$ or $u<0$. Now, in addition to its dependence on $g/\kappa$ and $\kappa T$,  the loading probability also depends on $\kappa T_0$, which represents the bandwidth of our downconversion process. In Fig.~\ref{MITNUloadpr}(a), we have plotted  the loading probability for different values of $g/\kappa$ with fixed values of $\kappa T$ and $\kappa T_0$ at $\gamma' =0$. Similar to what we found in the single-atom case, this figure shows that there exists an optimum value of $g/\kappa$ whose corresponding peak loading  probability is maximum. Hence, if we turn the control field off at this particular  time, the chance of finding both atoms in their excited states is maximum.  Figure~\ref{MITNUloadpr}(b) shows the same property from a different perspective. It plots the loading probability with $\kappa T$ and $g/ \kappa$ fixed for several values of $\kappa T_0$. Here, we see that there is an optimum value of $\kappa T_0$ for which the loading probability is maximum. 

\begin{figure}
\centering
\begin{tabular} {c}
\includegraphics[width=.7\linewidth]{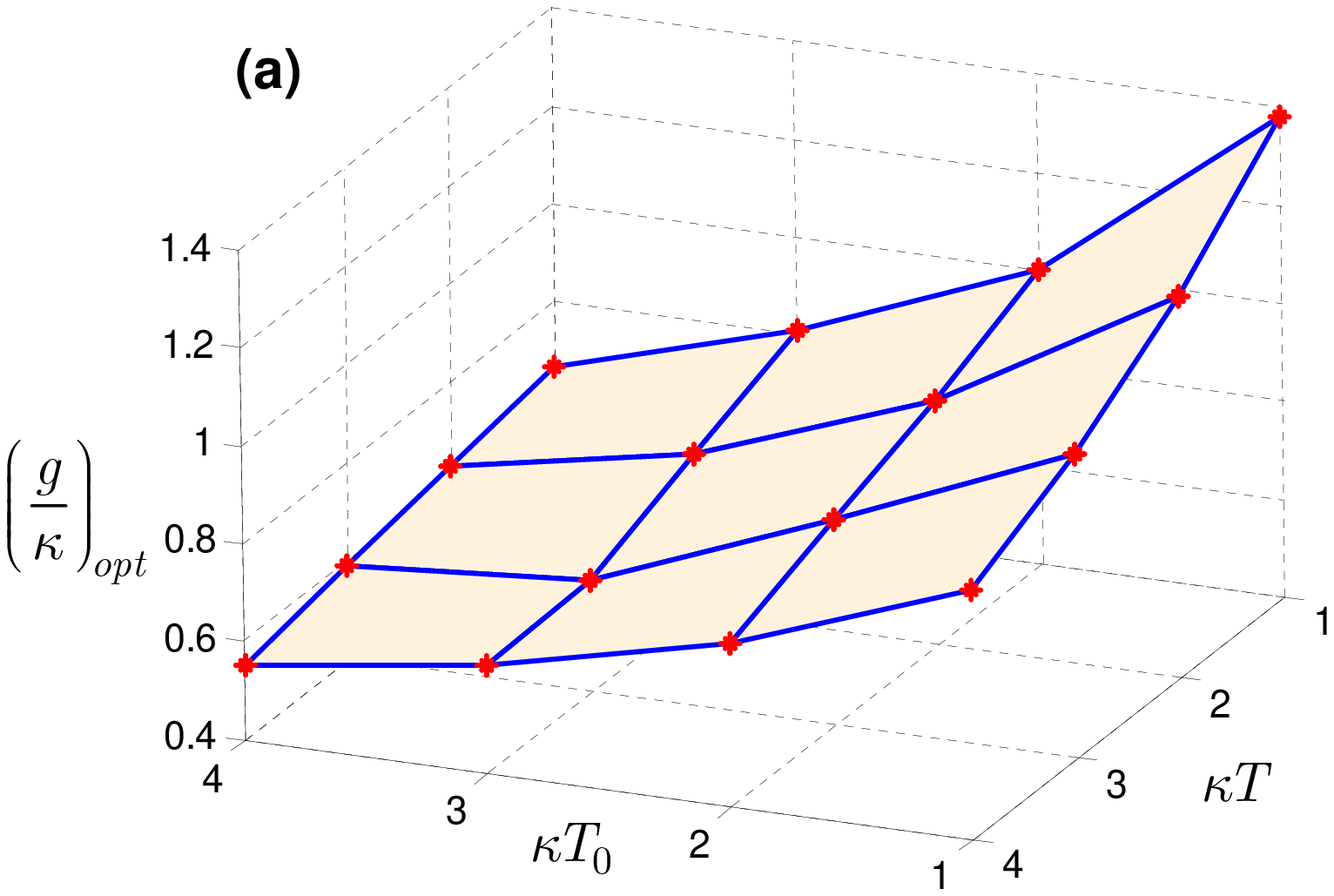} \\
\includegraphics[width=.7\linewidth]{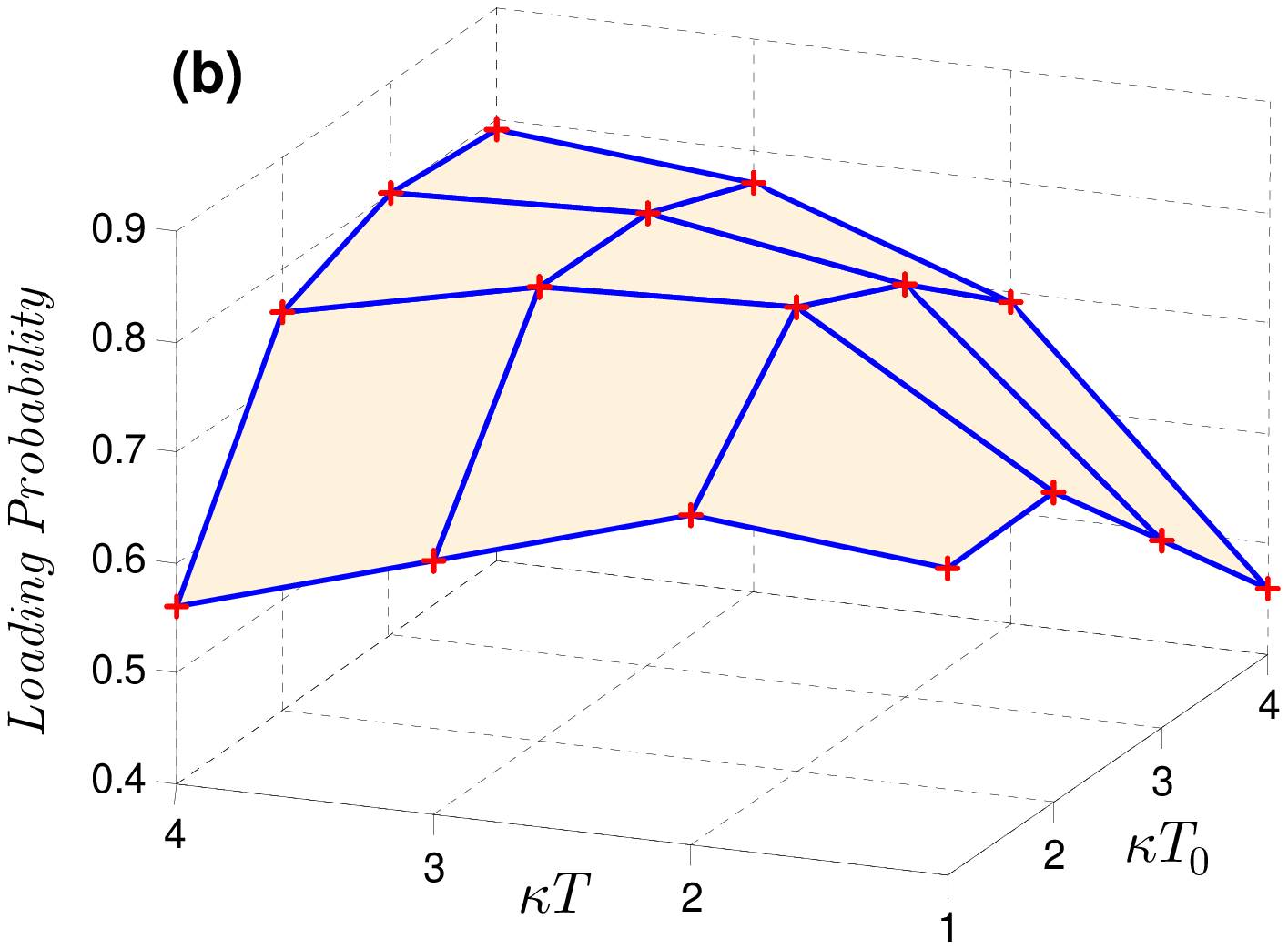} 
\end{tabular}
 \caption{ \label{MITNUdesign}
(Color online) (a) The optimum value of $g/\kappa$ versus $\kappa T$ and $\kappa T_0$ for the MIT-NU non-adiabatic loading problem. (b) The corresponding maximum loading probability at $(g/\kappa)_{opt}$. We have used a hyperbolic secant  pulse shape for the pump beam with $\gamma'=0$.}
\end{figure}

Figure~\ref{MITNUdesign} shows the optimum value of $g/\kappa$ and its corresponding loading probability as a function of $\kappa T$ and $\kappa T_0$ at $\gamma'=0$. From Fig.~\ref{MITNUdesign}(a), we observe that  higher values of $\kappa T$ or $\kappa T_0$, lead to lower values of $g/\kappa$. This is another  manifestation of the trade-off between the input pulse bandwidth and the required coupling rate. The lower the bandwidth is the lower the optimum coupling rate will be. Note that this regime of operation is not applicable to the adiabatic-passage approach on the two-photon resonance, in which the higher the coupling rate is the higher the loading probability will be; see Fig.~\ref{adbtwores}. As a result, the non-adiabatic loading mechanism has an advantage in that it allows us to use larger cavities and higher input bandwidths. The price,  however, is the value of loading probability that can be achieved. As shown in Fig.~\ref{MITNUdesign}(b), for non-adiabatic loading this probability is typically on the order of 70\%--80\%---rather than 100\%. From Fig.~\ref{MITNUdesign}(b), we see that higher loading probabilities occur for $\kappa T_0, \kappa T \geq 2$, where the cavity linewidth is wider than the bandwidths associated with the input pulse.

The initial plan for the MIT-NU architecture called for cavities with $\kappa \approx 5$~MHz and $g_c \approx 10$~MHz. That can be achieved with a cavity of length $L_c  \approx 400$~$\mu m$ \cite{trapatom1}, which corresponds to a Purcell factor $F_P = 3c/(4 \pi^2 \kappa L_c) \approx 4 \times 10^4$. That leaves a considerable flexibility for choosing $\Omega$ and $\Delta_1$ to operate at the optimum coupling rate $g=g_c \Omega /\Delta_1$. On the other hand, in order to have a high-throughput system it is required to have optical sources with narrow bandwidths comparable to $\kappa$. This can be achieved with the recent progress in building cavity-enhanced SPDCs \cite{chris}. Overall, although the loading probabilities that we have obtained in this section are under somewhat idealized conditions, their values are high enough to make the $5\,$dB fixed loss used in \cite{MIT/NU,Brent}, not an unduly optimistic, but a conceivably conservative assumption.

\section{CONCLUSIONS and discussion}
\label{SecConc}
Many proposals  for entanglement distribution over long distances rely on our ability to transfer the state of a single photon to an atomic quantum memory. Such a memory unit is comprised of either a single atom trapped in an optical cavity or an ensemble of atoms. In this paper, we addressed the problem of  loading a trapped atom with two-level, $\Lambda$-level, and double-$\Lambda$-level configurations.  The general approach suggested for solving this loading problem is based on adiabatic-transfer techniques. In this paper, we looked at the problem from a different viewpoint and realized that  a non-adiabatic mechanism may have characteristics that are comparable to, and in some aspects even better than, those of adiabatic loading. Our non-adiabatic method is based on the Rabi oscillation that naturally occurs when a single photon illuminates a trapped-atom quantum memory. We then freeze this oscillation when the probability of being in the desired state is maximum. We showed that by properly choosing the system parameters, e.g. atom-light coupling rate versus input bandwidth, we could achieve loading probabilities above 90\% in memory units with $\Lambda$ or double-$\Lambda$ atomic configurations. In terms of timing sensitivity, we showed that both adiabatic and non-adiabactic mechanisms have the same tolerance to errors in their estimate of photon's arrival time.

Whereas the adiabatic passage is inherently suited for long input pulses, our non-adiabatic technique does not impose any restrictions on the source bandwidth and therefore allows a larger class of optical sources to be employed. That freedom can potentially increase the total system throughput for distributing entanglement. 

For any input photon bandwidth, we realized the existence of an optimum coupling rate, in our scheme, that maximized the loading probability. This optimum coupling rate slightly increased when we introduced spontaneous decay, but it nevertheless remained in a range achievable by current technology. We also identified a trade-off between the required atom-light coupling strength and the input bandwidth, which showed that in order to accommodate a broadband source we needed to enhance our coupling rate. Conversely, the optimum coupling rate approached zero, although very slowly, as we decreased the bandwidth associated with the driving photon. We observed the existence of an optimum coupling rate in some adiabatic loading techniques as well. They  did not, however, offer such flexibility in working either with lower coupling rates or shorter input pulses.

Finally, we applied our initial results to the MIT-NU loading problem. The MIT-NU architecture uses a pair of trapped-rubidium-atom quantum memories, which are illuminated by the signal and idler outputs from an ultrabright doubly-resonant dual-OPA source of polarization-entangled photons. By approximating these outputs as a general biphoton state, we derived, analytically,  the loading probabilities for the non-adiabatic approach. In this case, we again observed the existence of an optimum coupling rate as a function of the driving pulse parameters. We showed that  loading probabilities above 80\% were achieveable at these optimum coupling rates, provided that  the input pulse's  bandwidth is narrower than that of optical cavities.

Although all our calculations are based on trapped-atom quantum memories, most of the results presented here can be extended to cold atomic-ensemble memories by including an additional multiplicative factor $\sqrt{N}$, with $N$ being the number of  atoms in the ensemble, in the coupling rate $g$. This factor accounts for the collective effect of ensemble. For room-temperature atomic ensembles, one should also account for the inhomogeneous broadening due to atomic motion, whose treatment needs a more general model than what we described in this paper.

\section*{ACKNOWLEDGMENTS} 
This work was supported by the MIT-HP Alliance.

\bibliography{main}

\begin{thebibliography}{43}
\expandafter\ifx\csname natexlab\endcsname\relax\def\natexlab#1{#1}\fi
\expandafter\ifx\csname bibnamefont\endcsname\relax
  \def\bibnamefont#1{#1}\fi
\expandafter\ifx\csname bibfnamefont\endcsname\relax
  \def\bibfnamefont#1{#1}\fi
\expandafter\ifx\csname citenamefont\endcsname\relax
  \def\citenamefont#1{#1}\fi
\expandafter\ifx\csname url\endcsname\relax
  \def\url#1{\texttt{#1}}\fi
\expandafter\ifx\csname urlprefix\endcsname\relax\def\urlprefix{URL }\fi
\providecommand{\bibinfo}[2]{#2}
\providecommand{\eprint}[2][]{\url{#2}}

\bibitem[{\citenamefont{Shor}(1994)}]{Shor94}
\bibinfo{author}{\bibfnamefont{P.~W.} \bibnamefont{Shor}}, in
  \emph{\bibinfo{booktitle}{Proc. 35th Symp. Found. of Comput. Sci., {IEEE}
  Press, Los Alamitos, CA}} (\bibinfo{year}{1994}).

\bibitem[{\citenamefont{Grover}(1996)}]{Gro96}
\bibinfo{author}{\bibfnamefont{L.}~\bibnamefont{Grover}}, in
  \emph{\bibinfo{booktitle}{Proc. 28th Annual {ACM} Symp. Theory of Comput.
  Sci.}} (\bibinfo{year}{1996}), p. \bibinfo{pages}{212}.

\bibitem[{\citenamefont{Bennett and Brassard}(1984)}]{BB84}
\bibinfo{author}{\bibfnamefont{C.~H.} \bibnamefont{Bennett}} \bibnamefont{and}
  \bibinfo{author}{\bibfnamefont{G.}~\bibnamefont{Brassard}}, in
  \emph{\bibinfo{booktitle}{Proc. of IEEE Int. Conf. on Comput. Syst. and
  Signal Process.}} (\bibinfo{year}{1984}), p. \bibinfo{pages}{175}.

\bibitem[{\citenamefont{Giovannetti et~al.}(2004)\citenamefont{Giovannetti,
  Lloyd, and Maccone}}]{prec_meas}
\bibinfo{author}{\bibfnamefont{V.}~\bibnamefont{Giovannetti}},
  \bibinfo{author}{\bibfnamefont{S.}~\bibnamefont{Lloyd}}, \bibnamefont{and}
  \bibinfo{author}{\bibfnamefont{L.}~\bibnamefont{Maccone}},
  \bibinfo{journal}{Science} \textbf{\bibinfo{volume}{306}},
  \bibinfo{pages}{1330} (\bibinfo{year}{2004}).

\bibitem[{com()}]{commercial}
\bibinfo{note}{See, for example, http://www.idquantique.com and
  http://www.magiqtech.com}.

\bibitem[{\citenamefont{Kuklewicz et~al.}(2006)\citenamefont{Kuklewicz, Wong,
  and Shapiro}}]{chris}
\bibinfo{author}{\bibfnamefont{C.~E.} \bibnamefont{Kuklewicz}},
  \bibinfo{author}{\bibfnamefont{F.~N.~C.} \bibnamefont{Wong}},
  \bibnamefont{and} \bibinfo{author}{\bibfnamefont{J.~H.}
  \bibnamefont{Shapiro}}, \textbf{\bibinfo{volume}{97}},
  \bibinfo{pages}{223601} (\bibinfo{year}{2006}).

\bibitem[{\citenamefont{Shapiro and Wong}(2000)}]{ultrabright}
\bibinfo{author}{\bibfnamefont{J.~H.} \bibnamefont{Shapiro}} \bibnamefont{and}
  \bibinfo{author}{\bibfnamefont{N.~C.} \bibnamefont{Wong}},
  \bibinfo{journal}{J. Opt. B: Quantum. and Semiclass. Opt.}
  \textbf{\bibinfo{volume}{2}}, \bibinfo{pages}{{L1}} (\bibinfo{year}{2000}).

\bibitem[{\citenamefont{Miller et~al.}(2003)\citenamefont{Miller, Nam,
  Martinis, and Sergienko}}]{SPD1}
\bibinfo{author}{\bibfnamefont{A.~J.} \bibnamefont{Miller}},
  \bibinfo{author}{\bibfnamefont{S.~W.} \bibnamefont{Nam}},
  \bibinfo{author}{\bibfnamefont{J.~M.} \bibnamefont{Martinis}},
  \bibnamefont{and} \bibinfo{author}{\bibfnamefont{A.~V.}
  \bibnamefont{Sergienko}}, \bibinfo{journal}{Appl. Phys. Lett.}
  \textbf{\bibinfo{volume}{83}}, \bibinfo{pages}{791} (\bibinfo{year}{2003}).

\bibitem[{\citenamefont{Rosfjord et~al.}(2006)\citenamefont{Rosfjord, Yang,
  Dauler, Kerman, Anant, Voronov, Gol'tsman, and Berggren}}]{SPD2}
\bibinfo{author}{\bibfnamefont{K.~M.} \bibnamefont{Rosfjord}},
  \bibinfo{author}{\bibfnamefont{J.~K.~W.} \bibnamefont{Yang}},
  \bibinfo{author}{\bibfnamefont{E.~A.} \bibnamefont{Dauler}},
  \bibinfo{author}{\bibfnamefont{A.~J.} \bibnamefont{Kerman}},
  \bibinfo{author}{\bibfnamefont{V.}~\bibnamefont{Anant}},
  \bibinfo{author}{\bibfnamefont{B.}~\bibnamefont{Voronov}},
  \bibinfo{author}{\bibfnamefont{G.~N.} \bibnamefont{Gol'tsman}},
  \bibnamefont{and} \bibinfo{author}{\bibfnamefont{K.~K.}
  \bibnamefont{Berggren}}, \bibinfo{journal}{Opt. Express}
  \textbf{\bibinfo{volume}{14}}, \bibinfo{pages}{527} (\bibinfo{year}{2006}).

\bibitem[{\citenamefont{Robinson et~al.}(2006)\citenamefont{Robinson, Kerman,
  Dauler, Barron, Caplan, Stevens, Carney, S.~A.~Hamilton, and
  Berggren}}]{SPD3}
\bibinfo{author}{\bibfnamefont{B.~S.} \bibnamefont{Robinson}},
  \bibinfo{author}{\bibfnamefont{A.~J.} \bibnamefont{Kerman}},
  \bibinfo{author}{\bibfnamefont{E.~A.} \bibnamefont{Dauler}},
  \bibinfo{author}{\bibfnamefont{R.~J.} \bibnamefont{Barron}},
  \bibinfo{author}{\bibfnamefont{D.~O.} \bibnamefont{Caplan}},
  \bibinfo{author}{\bibfnamefont{M.~L.} \bibnamefont{Stevens}},
  \bibinfo{author}{\bibfnamefont{J.~J.} \bibnamefont{Carney}},
  \bibinfo{author}{\bibfnamefont{J.~K. W.~Y.} \bibnamefont{S.~A.~Hamilton}},
  \bibnamefont{and} \bibinfo{author}{\bibfnamefont{K.~K.}
  \bibnamefont{Berggren}}, \bibinfo{journal}{Opt. Lett.}
  \textbf{\bibinfo{volume}{31}}, \bibinfo{pages}{444} (\bibinfo{year}{2006}).

\bibitem[{\citenamefont{Lloyd et~al.}(2001)\citenamefont{Lloyd, Shahriar,
  Shapiro, and Hemmer}}]{trap}
\bibinfo{author}{\bibfnamefont{S.}~\bibnamefont{Lloyd}},
  \bibinfo{author}{\bibfnamefont{M.~S.} \bibnamefont{Shahriar}},
  \bibinfo{author}{\bibfnamefont{J.~H.} \bibnamefont{Shapiro}},
  \bibnamefont{and} \bibinfo{author}{\bibfnamefont{P.~R.}
  \bibnamefont{Hemmer}}, \bibinfo{journal}{Phys. Rev. Lett.}
  \textbf{\bibinfo{volume}{87}}, \bibinfo{pages}{167903}
  (\bibinfo{year}{2001}).

\bibitem[{\citenamefont{Miller et~al.}(2005)\citenamefont{Miller, Northup,
  Birnbaum, Boca, Boozer, and Kimble}}]{trapatom1}
\bibinfo{author}{\bibfnamefont{R.}~\bibnamefont{Miller}},
  \bibinfo{author}{\bibfnamefont{T.~E.} \bibnamefont{Northup}},
  \bibinfo{author}{\bibfnamefont{K.~M.} \bibnamefont{Birnbaum}},
  \bibinfo{author}{\bibfnamefont{A.}~\bibnamefont{Boca}},
  \bibinfo{author}{\bibfnamefont{A.~D.} \bibnamefont{Boozer}},
  \bibnamefont{and} \bibinfo{author}{\bibfnamefont{H.~J.}
  \bibnamefont{Kimble}}, \bibinfo{journal}{J. Phys. B: Atomic, Molecular, and
  Optical Physics} \textbf{\bibinfo{volume}{38}}, \bibinfo{pages}{S551}
  (\bibinfo{year}{2005}).

\bibitem[{\citenamefont{Boca et~al.}(2004)\citenamefont{Boca, Miller, Birnbaum,
  Boozer, McKeever, and Kimble}}]{trapatom2}
\bibinfo{author}{\bibfnamefont{A.}~\bibnamefont{Boca}},
  \bibinfo{author}{\bibfnamefont{R.}~\bibnamefont{Miller}},
  \bibinfo{author}{\bibfnamefont{K.~M.} \bibnamefont{Birnbaum}},
  \bibinfo{author}{\bibfnamefont{A.~D.} \bibnamefont{Boozer}},
  \bibinfo{author}{\bibfnamefont{J.}~\bibnamefont{McKeever}}, \bibnamefont{and}
  \bibinfo{author}{\bibfnamefont{H.~J.} \bibnamefont{Kimble}},
  \bibinfo{journal}{Phys. Rev. Lett.} \textbf{\bibinfo{volume}{93}},
  \bibinfo{pages}{233603} (\bibinfo{year}{2004}).

\bibitem[{\citenamefont{McKeever et~al.}(2004)\citenamefont{McKeever, Boca,
  Boozer, Miller, Buck, Kuzmich, and Kimble}}]{trapatom3}
\bibinfo{author}{\bibfnamefont{J.}~\bibnamefont{McKeever}},
  \bibinfo{author}{\bibfnamefont{A.}~\bibnamefont{Boca}},
  \bibinfo{author}{\bibfnamefont{A.~D.} \bibnamefont{Boozer}},
  \bibinfo{author}{\bibfnamefont{R.}~\bibnamefont{Miller}},
  \bibinfo{author}{\bibfnamefont{J.~R.} \bibnamefont{Buck}},
  \bibinfo{author}{\bibfnamefont{A.}~\bibnamefont{Kuzmich}}, \bibnamefont{and}
  \bibinfo{author}{\bibfnamefont{H.~J.} \bibnamefont{Kimble}},
  \bibinfo{journal}{Science} \textbf{\bibinfo{volume}{303}},
  \bibinfo{pages}{1992} (\bibinfo{year}{2004}).

\bibitem[{\citenamefont{Duan et~al.}(2001)\citenamefont{Duan, Lukin, Cirac, and
  Zoller}}]{DLCZ}
\bibinfo{author}{\bibfnamefont{L.~M.} \bibnamefont{Duan}},
  \bibinfo{author}{\bibfnamefont{M.~D.} \bibnamefont{Lukin}},
  \bibinfo{author}{\bibfnamefont{J.~I.} \bibnamefont{Cirac}}, \bibnamefont{and}
  \bibinfo{author}{\bibfnamefont{P.}~\bibnamefont{Zoller}},
  \bibinfo{journal}{Nature} \textbf{\bibinfo{volume}{414}},
  \bibinfo{pages}{413} (\bibinfo{year}{2001}).

\bibitem[{\citenamefont{Chou et~al.}(2005)\citenamefont{Chou, de~Riedmatten,
  Felinto, Polyakov, van Enk, and Kimble}}]{cohtime2}
\bibinfo{author}{\bibfnamefont{C.~W.} \bibnamefont{Chou}},
  \bibinfo{author}{\bibfnamefont{H.}~\bibnamefont{de~Riedmatten}},
  \bibinfo{author}{\bibfnamefont{D.}~\bibnamefont{Felinto}},
  \bibinfo{author}{\bibfnamefont{S.~V.} \bibnamefont{Polyakov}},
  \bibinfo{author}{\bibfnamefont{S.~J.} \bibnamefont{van Enk}},
  \bibnamefont{and} \bibinfo{author}{\bibfnamefont{H.~J.}
  \bibnamefont{Kimble}}, \bibinfo{journal}{Nature}
  \textbf{\bibinfo{volume}{438}}, \bibinfo{pages}{828} (\bibinfo{year}{2005}).

\bibitem[{\citenamefont{{Chaneli\`{e}re}
  et~al.}(2005)\citenamefont{{Chaneli\`{e}re}, Matsukevich, Jenkins, Lan,
  Kennedy, and Kuzmich}}]{kuz2}
\bibinfo{author}{\bibfnamefont{T.}~\bibnamefont{{Chaneli\`{e}re}}},
  \bibinfo{author}{\bibfnamefont{D.~N.} \bibnamefont{Matsukevich}},
  \bibinfo{author}{\bibfnamefont{S.~D.} \bibnamefont{Jenkins}},
  \bibinfo{author}{\bibfnamefont{S.-Y.} \bibnamefont{Lan}},
  \bibinfo{author}{\bibfnamefont{T.~A.~B.} \bibnamefont{Kennedy}},
  \bibnamefont{and} \bibinfo{author}{\bibfnamefont{A.}~\bibnamefont{Kuzmich}},
  \bibinfo{journal}{Nature} \textbf{\bibinfo{volume}{438}},
  \bibinfo{pages}{833} (\bibinfo{year}{2005}).

\bibitem[{\citenamefont{{Chaneli\`{e}re}
  et~al.}(2006)\citenamefont{{Chaneli\`{e}re}, Matsukevich, Jenkins, Kennedy,
  Chapman, and Kuzmich}}]{Kuz1}
\bibinfo{author}{\bibfnamefont{T.}~\bibnamefont{{Chaneli\`{e}re}}},
  \bibinfo{author}{\bibfnamefont{D.~N.} \bibnamefont{Matsukevich}},
  \bibinfo{author}{\bibfnamefont{S.~D.} \bibnamefont{Jenkins}},
  \bibinfo{author}{\bibfnamefont{T.~A.~B.} \bibnamefont{Kennedy}},
  \bibinfo{author}{\bibfnamefont{M.~S.} \bibnamefont{Chapman}},
  \bibnamefont{and} \bibinfo{author}{\bibfnamefont{A.}~\bibnamefont{Kuzmich}},
  \bibinfo{journal}{Phys. Rev. Lett.} \textbf{\bibinfo{volume}{96}},
  \bibinfo{pages}{093604} (\bibinfo{year}{2006}).

\bibitem[{\citenamefont{Matsukevich et~al.}(2006)\citenamefont{Matsukevich,
  {Chaneli\`{e}re}, Jenkins, Lan, Kennedy, and Kuzmich}}]{Kuz3}
\bibinfo{author}{\bibfnamefont{D.~N.} \bibnamefont{Matsukevich}},
  \bibinfo{author}{\bibfnamefont{T.}~\bibnamefont{{Chaneli\`{e}re}}},
  \bibinfo{author}{\bibfnamefont{S.~D.} \bibnamefont{Jenkins}},
  \bibinfo{author}{\bibfnamefont{S.-Y.} \bibnamefont{Lan}},
  \bibinfo{author}{\bibfnamefont{T.~A.~B.} \bibnamefont{Kennedy}},
  \bibnamefont{and} \bibinfo{author}{\bibfnamefont{A.}~\bibnamefont{Kuzmich}},
  \bibinfo{journal}{Phys. Rev. Lett.} \textbf{\bibinfo{volume}{96}},
  \bibinfo{pages}{030405} (\bibinfo{year}{2006}).

\bibitem[{\citenamefont{de~Riedmatten et~al.}(2006)\citenamefont{de~Riedmatten,
  Laurat, Chou, Schomburg, Felinto, and Kimble}}]{cohtime3}
\bibinfo{author}{\bibfnamefont{H.}~\bibnamefont{de~Riedmatten}},
  \bibinfo{author}{\bibfnamefont{J.}~\bibnamefont{Laurat}},
  \bibinfo{author}{\bibfnamefont{C.~W.} \bibnamefont{Chou}},
  \bibinfo{author}{\bibfnamefont{E.~W.} \bibnamefont{Schomburg}},
  \bibinfo{author}{\bibfnamefont{D.}~\bibnamefont{Felinto}}, \bibnamefont{and}
  \bibinfo{author}{\bibfnamefont{H.~J.} \bibnamefont{Kimble}},
  \bibinfo{journal}{Phys. Rev. Lett.} \textbf{\bibinfo{volume}{97}},
  \bibinfo{pages}{113603} (\bibinfo{year}{2006}).

\bibitem[{\citenamefont{Bennett et~al.}(1993)\citenamefont{Bennett, Brassard,
  {Cr\'{e}peau}, Jozsa, Peres, and Wootters}}]{Bennett}
\bibinfo{author}{\bibfnamefont{C.~H.} \bibnamefont{Bennett}},
  \bibinfo{author}{\bibfnamefont{G.}~\bibnamefont{Brassard}},
  \bibinfo{author}{\bibfnamefont{C.}~\bibnamefont{{Cr\'{e}peau}}},
  \bibinfo{author}{\bibfnamefont{R.}~\bibnamefont{Jozsa}},
  \bibinfo{author}{\bibfnamefont{A.}~\bibnamefont{Peres}}, \bibnamefont{and}
  \bibinfo{author}{\bibfnamefont{W.~K.} \bibnamefont{Wootters}},
  \bibinfo{journal}{Phys. Rev. Lett.} \textbf{\bibinfo{volume}{70}},
  \bibinfo{pages}{1895} (\bibinfo{year}{1993}).

\bibitem[{\citenamefont{Chen et~al.}(2002)\citenamefont{Chen, Hogg, and
  Beausoleil}}]{Ray1}
\bibinfo{author}{\bibfnamefont{K.-Y.} \bibnamefont{Chen}},
  \bibinfo{author}{\bibfnamefont{T.}~\bibnamefont{Hogg}}, \bibnamefont{and}
  \bibinfo{author}{\bibfnamefont{R.}~\bibnamefont{Beausoleil}},
  \bibinfo{journal}{Quantum Inf. Process.} \textbf{\bibinfo{volume}{1}},
  \bibinfo{pages}{449} (\bibinfo{year}{2002}).

\bibitem[{\citenamefont{Shapiro}(2002)}]{MIT/NU}
\bibinfo{author}{\bibfnamefont{J.~H.} \bibnamefont{Shapiro}},
  \bibinfo{journal}{New J. Phys.} \textbf{\bibinfo{volume}{4}},
  \bibinfo{pages}{art.~47} (\bibinfo{year}{2002}).

\bibitem[{\citenamefont{Bose et~al.}(1998)\citenamefont{Bose, Vedral, and
  Knight}}]{swap}
\bibinfo{author}{\bibfnamefont{S.}~\bibnamefont{Bose}},
  \bibinfo{author}{\bibfnamefont{V.}~\bibnamefont{Vedral}}, \bibnamefont{and}
  \bibinfo{author}{\bibfnamefont{P.~L.} \bibnamefont{Knight}},
  \bibinfo{journal}{Phys. Rev. A} \textbf{\bibinfo{volume}{57}},
  \bibinfo{pages}{822} (\bibinfo{year}{1998}).

\bibitem[{\citenamefont{Duan and Kimble}(2003)}]{DK03}
\bibinfo{author}{\bibfnamefont{L.~M.} \bibnamefont{Duan}} \bibnamefont{and}
  \bibinfo{author}{\bibfnamefont{H.~J.} \bibnamefont{Kimble}},
  \bibinfo{journal}{Phys. Rev. Lett.} \textbf{\bibinfo{volume}{90}},
  \bibinfo{pages}{253601} (\bibinfo{year}{2003}).

\bibitem[{\citenamefont{Cirac et~al.}(1997)\citenamefont{Cirac, Zoller, Kimble,
  and Mabuchi}}]{nontelep}
\bibinfo{author}{\bibfnamefont{J.~I.} \bibnamefont{Cirac}},
  \bibinfo{author}{\bibfnamefont{P.}~\bibnamefont{Zoller}},
  \bibinfo{author}{\bibfnamefont{H.~J.} \bibnamefont{Kimble}},
  \bibnamefont{and} \bibinfo{author}{\bibfnamefont{H.}~\bibnamefont{Mabuchi}},
  \bibinfo{journal}{Phys. Rev. Lett.} \textbf{\bibinfo{volume}{78}},
  \bibinfo{pages}{3221} (\bibinfo{year}{1997}).

\bibitem[{\citenamefont{Kwiat et~al.}(1995)\citenamefont{Kwiat, Mattle,
  Weinfurter, Zeilinger, Sergienko, and Shih}}]{PDC1}
\bibinfo{author}{\bibfnamefont{P.~G.} \bibnamefont{Kwiat}},
  \bibinfo{author}{\bibfnamefont{K.}~\bibnamefont{Mattle}},
  \bibinfo{author}{\bibfnamefont{H.}~\bibnamefont{Weinfurter}},
  \bibinfo{author}{\bibfnamefont{A.}~\bibnamefont{Zeilinger}},
  \bibinfo{author}{\bibfnamefont{A.~V.} \bibnamefont{Sergienko}},
  \bibnamefont{and} \bibinfo{author}{\bibfnamefont{Y.}~\bibnamefont{Shih}},
  \bibinfo{journal}{Phys. Rev. Lett.} \textbf{\bibinfo{volume}{75}},
  \bibinfo{pages}{4337} (\bibinfo{year}{1995}).

\bibitem[{\citenamefont{Kwiat et~al.}(1999)\citenamefont{Kwiat, Waks, White,
  Appelbaum, and Eberhard}}]{PDC2}
\bibinfo{author}{\bibfnamefont{P.~G.} \bibnamefont{Kwiat}},
  \bibinfo{author}{\bibfnamefont{E.}~\bibnamefont{Waks}},
  \bibinfo{author}{\bibfnamefont{A.~G.} \bibnamefont{White}},
  \bibinfo{author}{\bibfnamefont{I.}~\bibnamefont{Appelbaum}},
  \bibnamefont{and} \bibinfo{author}{\bibfnamefont{P.~H.}
  \bibnamefont{Eberhard}}, \bibinfo{journal}{Phys. Rev. A}
  \textbf{\bibinfo{volume}{60}}, \bibinfo{pages}{R773} (\bibinfo{year}{1999}).

\bibitem[{\citenamefont{{K\"onig} et~al.}(2005)\citenamefont{{K\"onig}, Mason,
  Wong, and Albota}}]{PDC3}
\bibinfo{author}{\bibfnamefont{F.}~\bibnamefont{{K\"onig}}},
  \bibinfo{author}{\bibfnamefont{E.~J.} \bibnamefont{Mason}},
  \bibinfo{author}{\bibfnamefont{F.~N.~C.} \bibnamefont{Wong}},
  \bibnamefont{and} \bibinfo{author}{\bibfnamefont{M.~A.}
  \bibnamefont{Albota}}, \bibinfo{journal}{Phys. Rev. A}
  \textbf{\bibinfo{volume}{71}}, \bibinfo{pages}{033805}
  (\bibinfo{year}{2005}).

\bibitem[{\citenamefont{Fiorentino et~al.}(2005)\citenamefont{Fiorentino,
  Kuklewicz, and Wong}}]{PDC4}
\bibinfo{author}{\bibfnamefont{M.}~\bibnamefont{Fiorentino}},
  \bibinfo{author}{\bibfnamefont{C.}~\bibnamefont{Kuklewicz}},
  \bibnamefont{and} \bibinfo{author}{\bibfnamefont{F.~N.~C.}
  \bibnamefont{Wong}}, \bibinfo{journal}{Opt. Express}
  \textbf{\bibinfo{volume}{13}}, \bibinfo{pages}{127} (\bibinfo{year}{2005}).

\bibitem[{\citenamefont{Kuhr et~al.}(2005)\citenamefont{Kuhr, Alt, Schrader,
  Dotsenko, Miroshnychenko, Rauschenbeutel, and Meschede}}]{dephasing}
\bibinfo{author}{\bibfnamefont{S.}~\bibnamefont{Kuhr}},
  \bibinfo{author}{\bibfnamefont{W.}~\bibnamefont{Alt}},
  \bibinfo{author}{\bibfnamefont{D.}~\bibnamefont{Schrader}},
  \bibinfo{author}{\bibfnamefont{I.}~\bibnamefont{Dotsenko}},
  \bibinfo{author}{\bibfnamefont{Y.}~\bibnamefont{Miroshnychenko}},
  \bibinfo{author}{\bibfnamefont{A.}~\bibnamefont{Rauschenbeutel}},
  \bibnamefont{and} \bibinfo{author}{\bibfnamefont{D.}~\bibnamefont{Meschede}},
  \bibinfo{journal}{Phys. Rev. A} \textbf{\bibinfo{volume}{72}},
  \bibinfo{pages}{023406} (\bibinfo{year}{2005}).

\bibitem[{\citenamefont{Felinto et~al.}(2005)\citenamefont{Felinto, Chou,
  de~Riedmatten, Polyakov, and Kimble}}]{cohtime}
\bibinfo{author}{\bibfnamefont{D.}~\bibnamefont{Felinto}},
  \bibinfo{author}{\bibfnamefont{C.~W.} \bibnamefont{Chou}},
  \bibinfo{author}{\bibfnamefont{H.}~\bibnamefont{de~Riedmatten}},
  \bibinfo{author}{\bibfnamefont{S.~V.} \bibnamefont{Polyakov}},
  \bibnamefont{and} \bibinfo{author}{\bibfnamefont{H.~J.}
  \bibnamefont{Kimble}}, \bibinfo{journal}{Phys. Rev. A}
  \textbf{\bibinfo{volume}{72}}, \bibinfo{pages}{053809}
  (\bibinfo{year}{2005}).

\bibitem[{\citenamefont{Fleischhauer et~al.}(2000)\citenamefont{Fleischhauer,
  Yelin, and Lukin}}]{fleisch}
\bibinfo{author}{\bibfnamefont{M.}~\bibnamefont{Fleischhauer}},
  \bibinfo{author}{\bibfnamefont{S.~F.} \bibnamefont{Yelin}}, \bibnamefont{and}
  \bibinfo{author}{\bibfnamefont{M.~D.} \bibnamefont{Lukin}},
  \bibinfo{journal}{Opt. Commun.} \textbf{\bibinfo{volume}{179}},
  \bibinfo{pages}{395} (\bibinfo{year}{2000}).

\bibitem[{\citenamefont{Yen and Shapiro}(2003)}]{Brent}
\bibinfo{author}{\bibfnamefont{B.~J.} \bibnamefont{Yen}} \bibnamefont{and}
  \bibinfo{author}{\bibfnamefont{J.~H.} \bibnamefont{Shapiro}},
  \bibinfo{journal}{{IEEE} J. Sel. Topics Quantum Electron.}
  \textbf{\bibinfo{volume}{9}}, \bibinfo{pages}{1483} (\bibinfo{year}{2003}).

\bibitem[{\citenamefont{Gardiner and Collett}(1985)}]{GC}
\bibinfo{author}{\bibfnamefont{C.~W.} \bibnamefont{Gardiner}} \bibnamefont{and}
  \bibinfo{author}{\bibfnamefont{M.~J.} \bibnamefont{Collett}},
  \bibinfo{journal}{Phys. Rev. A} \textbf{\bibinfo{volume}{31}},
  \bibinfo{pages}{3761} (\bibinfo{year}{1985}).

\bibitem[{\citenamefont{Blow et~al.}(1990)\citenamefont{Blow, Loudon, Phoenix,
  and Shepherd}}]{Loudon}
\bibinfo{author}{\bibfnamefont{K.~J.} \bibnamefont{Blow}},
  \bibinfo{author}{\bibfnamefont{R.}~\bibnamefont{Loudon}},
  \bibinfo{author}{\bibfnamefont{S.~J.~D.} \bibnamefont{Phoenix}},
  \bibnamefont{and} \bibinfo{author}{\bibfnamefont{T.~J.}
  \bibnamefont{Shepherd}}, \bibinfo{journal}{Phys. Rev. A}
  \textbf{\bibinfo{volume}{42}}, \bibinfo{pages}{4102} (\bibinfo{year}{1990}).

\bibitem[{\citenamefont{Scully and Zubairy}(1997)}]{SZ}
\bibinfo{author}{\bibfnamefont{M.~O.} \bibnamefont{Scully}} \bibnamefont{and}
  \bibinfo{author}{\bibfnamefont{M.~S.} \bibnamefont{Zubairy}},
  \emph{\bibinfo{title}{Quantum Optics}} (\bibinfo{publisher}{Cambridge
  University Press}, \bibinfo{address}{New York}, \bibinfo{year}{1997}).

\bibitem[{\citenamefont{Razavi et~al.}(2006)\citenamefont{Razavi, Giovannetti,
  Maccone, and Shapiro}}]{QELS}
\bibinfo{author}{\bibfnamefont{M.}~\bibnamefont{Razavi}},
  \bibinfo{author}{\bibfnamefont{V.}~\bibnamefont{Giovannetti}},
  \bibinfo{author}{\bibfnamefont{L.}~\bibnamefont{Maccone}}, \bibnamefont{and}
  \bibinfo{author}{\bibfnamefont{J.~H.} \bibnamefont{Shapiro}}, in
  \emph{\bibinfo{booktitle}{{Technical Digest}, QFA7, Quantum Electron. and
  Laser Sci. Conf.}} (\bibinfo{year}{2006}).

\bibitem[{\citenamefont{Kuklinski et~al.}(1989)\citenamefont{Kuklinski,
  Gaubatz, Hioe, and Bergmann}}]{stirap}
\bibinfo{author}{\bibfnamefont{J.~R.} \bibnamefont{Kuklinski}},
  \bibinfo{author}{\bibfnamefont{U.}~\bibnamefont{Gaubatz}},
  \bibinfo{author}{\bibfnamefont{F.~T.} \bibnamefont{Hioe}}, \bibnamefont{and}
  \bibinfo{author}{\bibfnamefont{K.}~\bibnamefont{Bergmann}},
  \bibinfo{journal}{Phys. Rev. A} \textbf{\bibinfo{volume}{40}},
  \bibinfo{pages}{6741} (\bibinfo{year}{1989}).

\bibitem[{\citenamefont{Parkins et~al.}(1995)\citenamefont{Parkins, Marte,
  Zoller, Carnal, and Kimble}}]{kimble_adiab}
\bibinfo{author}{\bibfnamefont{A.~S.} \bibnamefont{Parkins}},
  \bibinfo{author}{\bibfnamefont{P.}~\bibnamefont{Marte}},
  \bibinfo{author}{\bibfnamefont{P.}~\bibnamefont{Zoller}},
  \bibinfo{author}{\bibfnamefont{O.}~\bibnamefont{Carnal}}, \bibnamefont{and}
  \bibinfo{author}{\bibfnamefont{H.~J.} \bibnamefont{Kimble}},
  \bibinfo{journal}{Phys. Rev. A} \textbf{\bibinfo{volume}{51}},
  \bibinfo{pages}{1578} (\bibinfo{year}{1995}).

\bibitem[{\citenamefont{Lloyd et~al.}(2004)\citenamefont{Lloyd, Shapiro, Wong,
  Kumar, Shahriar, and Yuen}}]{ACM}
\bibinfo{author}{\bibfnamefont{S.}~\bibnamefont{Lloyd}},
  \bibinfo{author}{\bibfnamefont{J.~H.} \bibnamefont{Shapiro}},
  \bibinfo{author}{\bibfnamefont{F.~N.~C.} \bibnamefont{Wong}},
  \bibinfo{author}{\bibfnamefont{P.}~\bibnamefont{Kumar}},
  \bibinfo{author}{\bibfnamefont{S.~M.} \bibnamefont{Shahriar}},
  \bibnamefont{and} \bibinfo{author}{\bibfnamefont{H.~P.} \bibnamefont{Yuen}},
  \bibinfo{journal}{Computer Commun. Rev.} \textbf{\bibinfo{volume}{34}},
  \bibinfo{pages}{9} (\bibinfo{year}{2004}).

\bibitem[{\citenamefont{Albota et~al.}(2006)\citenamefont{Albota, Wong, and
  Shapiro}}]{Albota}
\bibinfo{author}{\bibfnamefont{M.~A.} \bibnamefont{Albota}},
  \bibinfo{author}{\bibfnamefont{F.~N.~C.} \bibnamefont{Wong}},
  \bibnamefont{and} \bibinfo{author}{\bibfnamefont{J.~H.}
  \bibnamefont{Shapiro}}, \bibinfo{journal}{J. Opt. Soc. Am. B}
  \textbf{\bibinfo{volume}{23}}, \bibinfo{pages}{918} (\bibinfo{year}{2006}).

\bibitem[{\citenamefont{Giovannetti et~al.}(2002)\citenamefont{Giovannetti,
  Maccone, Shapiro, and Wong}}]{TBDB}
\bibinfo{author}{\bibfnamefont{V.}~\bibnamefont{Giovannetti}},
  \bibinfo{author}{\bibfnamefont{L.}~\bibnamefont{Maccone}},
  \bibinfo{author}{\bibfnamefont{J.~H.} \bibnamefont{Shapiro}},
  \bibnamefont{and} \bibinfo{author}{\bibfnamefont{F.~N.~C.}
  \bibnamefont{Wong}}, \bibinfo{journal}{Phys. Rev. Lett.}
  \textbf{\bibinfo{volume}{88}}, \bibinfo{pages}{183602}
  (\bibinfo{year}{2002}).

\end{thebibliography}

\end{document}